\documentclass[11pt,a4paper]{article}
\usepackage{jheppub}
\usepackage{graphicx}
\usepackage{dcolumn}
\usepackage{bm}
\usepackage{color}
\usepackage{xcolor}
\usepackage{soul}
\usepackage[normalem]{ulem}

\usepackage{slashed,physics}
\usepackage{tikz}
\usepackage{tikz-feynman}
\usepackage{multirow}
\usepackage[capitalize]{cleveref}
\usepackage{booktabs}
\usepackage{hyperref}
\usepackage{comment}

\newcommand{\be}{\begin{eqnarray}}
\newcommand{\ee}{\end{eqnarray}}
\newcommand{\beq}{\begin{equation}}
\newcommand{\eeq}{\end{equation}}

\newcommand{\bea}{\begin{eqnarray}}
\newcommand{\eea}{\end{eqnarray}}
\newcommand{\bear}{\begin{eqnarray}}
\newcommand{\eear}{\end{eqnarray}}
\newcommand{\ba}{\begin{array}}
\newcommand{\ea}{\end{array}}

\newcommand{\bench}[1]{{\color{black}#1}}

\newcommand{\SUp}{\ensuremath{\text{SU}(15)_\text{p}}}
\newcommand{\SUF}{\ensuremath{\text{SU}(4)_\text{F}}}
\newcommand{\SUI}{\ensuremath{\text{SU}(2)_\text{I}}}

\preprint{CERN-TH-2026-037}
\title{The flavour of $SU(15)$ composite quarks and leptons}

\author[a]{Beno\^it Assi,}
\author[b]{Ryan Plestid,}
\author[c,d]{Amartya Sengupta,}
\author[a]{Jure Zupan}

\affiliation[a]{Department of Physics, University of Cincinnati, Cincinnati, Ohio 45221, USA}
\affiliation[b]{Theoretical Physics Department, CERN, 1 Esplanade des Particules, CH-1211 Geneva 23, Switzerland}
\affiliation[c]{Department of Physics, The State University of New York, Buffalo, New York 14260, USA}
\affiliation[d]{International Center for Quantum-field Measurement Systems for Studies of the Universe and Particles (QUP),
KEK, 1-1 Oho, Tsukuba, Ibaraki 305-0801, Japan}

\emailAdd{assibt@ucmail.uc.edu}
\emailAdd{ryan.plestid@cern.ch}
\emailAdd{amartyas@buffalo.edu}
\emailAdd{zupanje@ucmail.uc.edu}

\abstract{ We study the flavour structure of an $\SUp$ confining chiral gauge theory in which the Standard Model (SM) quarks, leptons, and Higgs emerge as composite bound states. The couplings of two scalar fields in the conjugate antisymmetric ($\overline{\mathbf{105}}$) and conjugate symmetric ($\overline{\mathbf{120}}$) representations of $\SUp$ provide two $\SUF$ flavour-breaking spurions that generate both the SM Yukawa couplings and the flavour-changing processes. The up and down Yukawa matrices are tightly-correlated due to a ``right-handed isospin'' symmetry, which predicts a trivial CKM matrix in the absence of spontaneous symmetry breaking. The lepton Yukawas are correlated with the quarks due to a common source of flavour spurions. With a judicious Froggatt--Nielsen--like texture for the two flavour spurions  we find that a benchmark fit with $\mathcal{O}(1)$ non-perturbative coefficients reproduces all six quark masses, three charged lepton masses, and the CKM matrix. The same spurions  mediate charged lepton flavour violation, neutral meson mixing, rare kaon decays, and induce electric dipole moments. We compare the reach on the compositeness scale $\Lambda_{\rm pre}$ across these observables in the numerical benchmark and find that the electron EDM and $K^0$-$\bar{K}^0$ mixing provide the strongest sensitivity, reaching ${\mathcal O}(10^4)\,$TeV, the lower end of the range probed by proton decay, while $\mu\to e\gamma$, $D^0$-$\bar{D}^0$ mixing, and $\mu$-$e$ conversion give complementary reach at $10^2$--$10^3\,$TeV. The projected electron EDM sensitivity extends this to ${\mathcal O}(10^6)\,$TeV, beyond the reach of planned proton decay searches. 
}

\keywords{Beyond Standard Model, Technicolour and Composite Models, Quark Masses and SM Parameters}

\begin{document}

\maketitle

\pagebreak 

\section{Introduction}
\label{sec:intro}

Atoms, nuclei, and nucleons are all composed of more elementary constituents: quarks, gluons, and leptons. The composite nature of atoms, nuclei, and nucleons reveals itself as one performs experiments at higher and higher energies, thereby probing shorter and shorter distances. While all current experimental evidence is consistent with quarks and leptons being elementary, the possibility that quarks and leptons are composite, being composed of ``preon'' bound states characterized by a length scale below the resolution of the LHC, remains a compelling possibility. 

This idea has a long history \cite{Peskin:1981ak}. The preon models initially faced the challenge that the masses of the composite particles were comparable to the confinement scale \cite{Pati:1975md,Terazawa:1979pj,Shupe:1979fv,Harari:1979gi}; this is clearly at odds with experimental data. A mechanism for generating light fermionic composite bound states was realized based on chiral gauge theories \cite{Dimopoulos:1980hn,Raby:1979my} which must be anomaly free and satisfy a set of non-trivial 't Hooft anomaly matching constraints \cite{tHooft:1979rat} and can be further constrained with large-$N$ analysis \cite{Eichten:1985fs}. Composite dynamics have also been proposed as an explanation of the Standard Model's flavour structure via a mixing between an elementary and composite sectors (partial compositeness) \cite{Kaplan:1991dc,Panico:2015jxa}. Despite these theoretical developments, a central difficulty that has persisted is to find models that could host the Standard Model as an emergent effective theory in the infrared.

A renormalizable preon model whose dynamics can plausibly give rise to the Standard Model (SM) has recently been proposed in Ref.~\cite{Dobrescu:2021fny} and further developed in~\cite{Assi:2022jwg,Assi:2025rjx}. The confinement mechanism is an $\SUp$ gauge theory where quarks, leptons, and the Higgs all arise as composite bound states. The model contains $19$ chiral fermions in the fundamental representation (of which $15$ carry SM gauge charges and $4$ are singlets) and one fermion in the conjugate symmetric representation; this field content is required by anomaly cancellation and is expected to confine based on large-$N$ arguments~\cite{Eichten:1985fs}. Furthermore, the model satisfies non-trivial 't Hooft anomaly matching conditions that suggest it hosts chiral fermion bound states. The four singlet preons $\psi_i$ furnish an $\SUF$ global flavour symmetry that is spontaneously broken to $\text{SU(3)}_\text{F}$ by a composite scalar vacuum expectation value (vev), yielding three SM generations. Large-$N$ counting and symmetry structures naturally suppress baryon number violation~\cite{Assi:2022jwg}, so that the compositeness scale can be as low as $\Lambda_{\rm pre} \gtrsim 10^5$~TeV while remaining consistent with proton decay constraints. 
This relatively accessible scale opens the door to tests using precision flavour physics, which is the focus of the present manuscript.

Realistic fermion masses and CKM mixing require explicit breaking of $\SUF$. The simplest option, studied here, is a pair of scalar fields $\mathcal{A}$ and $\mathcal{A}'$ transforming in the conjugate antisymmetric $\overline{\mathbf{105}}$ and symmetric $\overline{\mathbf{120}}$ representations of $\SUp$, respectively.\footnote{Previous studies assumed both to be in a $\overline{\mathbf{105}}$ representation. Upon closer inspection, the Yukawa couplings used in fact correspond to the $\overline{\mathbf{120}}$ instead; we correct this minor oversight here.} The corresponding Yukawa couplings to the singlet preons, $\lambda_{ij}$ and $\lambda'_{ij}$, serve as $\SUF$ flavour spurions~\cite{Dobrescu:2021fny,Assi:2022jwg,Assi:2025rjx}. Two matrices are needed because one of the two can always be diagonalized by a field redefinition; the second provides the misalignment necessary for a non-trivial CKM matrix. A realistic CKM furthermore requires the two scalars to be in different representations of $\SUp$, as we show below. These same spurions also mediate flavour-changing neutral current (FCNC) processes via scalar exchanges at the scale $\Lambda_{\rm pre}$. The Yukawa coupling matrices therefore simultaneously control the pattern of SM fermion masses and of flavour violation. This tight connection between masses and FCNCs makes the model predictive and amenable to experimental tests, despite non-perturbative $\SUp$ dynamics.

The purpose of this paper is twofold: first, to determine whether the model can simultaneously accommodate the observed fermion mass hierarchies and CKM mixing; and second, to assess which flavour observables can be used as sensitive probes of dynamics at $\Lambda_{\rm pre}$. Previous work on this model~\cite{Dobrescu:2021fny,Assi:2022jwg} focused primarily on baryon number violation and the composite vectorlike fermion spectrum, leaving the detailed flavour structure largely unexplored. Here we construct the quark and charged lepton Yukawa matrices\footnote{Since the structure of neutrino masses may be sensitive to UV physics well above $\Lambda_{\rm pre}$, the model building implications of the observed neutrino mass matrices are left for future work.} at first nontrivial order in $1/N$, where $N=15$ is the number of $\SUp$ colours, parametrizing the non-perturbative $\SUp$ dynamics through a finite number of real coefficients whose size we estimate using naive dimensional analysis (NDA). We impose a hierarchical texture on the flavour-breaking spurions $\lambda$, $\tilde\lambda$, controlled by a small expansion parameter $\kappa \sim$~$0.17$ that seeds inter-generation mass ratios spanning five orders of magnitude, and identify a benchmark fit that reproduces all six quark masses, three charged lepton masses, and the CKM matrix.

With the benchmark coupling matrices in hand, we evaluate the most important flavour probes of $\Lambda_{\rm pre}$. These come from muon experiments ($\mu \to e\gamma$, coherent $\mu$-$e$ conversion in nuclei, and $\mu\to 3e$ \cite{Davidson:2022nnl}), kaon physics ($K\rightarrow \pi \nu\bar\nu$, and $K^0$-$\bar{K}^0$ mixing \cite{Aebischer:2025mwl,Buras:2018wmb}) and the electron electric dipole moment ($e$-EDM) \cite{Pospelov:2005pr,Pospelov:2025vzj}. We find that electron EDM and $K^0$-$\bar{K}^0$ mixing through $\epsilon_K$ provide the strongest constraints, both around $\Lambda_{\rm pre}\sim 10^4$\,TeV, near the proton decay reach, while $\mu\to e\gamma$, $D^0$-$\bar D^0$ mixing, and dipole-mediated $\mu$-$e$ conversion give complementary reach in the $10^2$--$10^3$~TeV range.

The paper is organized as follows. In \cref{sec:model} we review the preon model, its field content, and the large-$N$ power counting rules. 
The quark and charged lepton Yukawa matrices are constructed in \cref{sec:masses}, where we also present the benchmark fit. The phenomenological analysis of flavour-changing processes is carried out in \cref{sec:flavour}, and a comparative summary of the experimental reach is given in \cref{sec:results-discussion}. We conclude in \cref{sec:conclusions}. Appendix \ref{app:sec:spurions} contains further details about flavour breaking spurions, while appendices \ref{app:parameters} and \ref{app:minimization} contain the full set of fitted parameters and details of the numerical minimization, respectively. 

\section{The two scalar preon model}
\label{sec:model}

\begin{table}[t]
    \centering
    \renewcommand{\arraystretch}{1.4}
    \begin{tabular}{cccc}
    \hline\hline
         Field & Spin & $\SUp$ & $\text{SU(3)}_\text{c} \times \text{SU(2)}_\text{L} \times \text{U(1)}_\text{Y}$ \\
         \hline
         $\psi_Q$ & 1/2 & $\Box$ & $(3, 2, +1/6)$ \\
         $\psi_U$ & 1/2 & $\Box$ & $(\bar{3}, 1, -2/3)$ \\
         $\psi_D$ & 1/2 & $\Box$ & $(\bar{3}, 1, +1/3)$ \\
         $\psi_L$ & 1/2 & $\Box$ & $(1, 2, -1/2)$ \\
         $\psi_E$ & 1/2 & $\Box$ & $(1, 1, +1)$ \\
         $\psi_1, \psi_2, \psi_3, \psi_4$ & 1/2 & $\Box$ & $(1, 1, 0)$ \\
         $\Omega$ & 1/2 & $\overset{ \overset{ \rule{3.2ex}{0.07em} }{ {} } }{ \Box\!\Box } $ & $(1, 1, 0)$ \\
         \hline \\[-20pt]
         $\cal A$ & 0 & $\overset{ \overset{ \rule{1.6ex}{0.07em} }{ {} } }{ \parbox[c]{0.3cm}{$\Box$ \\[-2.95mm] $\Box$ \\ [-4.5mm] } }$ & $(1, 1, 0)$ \\
         $\cal A'$ & 0 & $\overset{ \overset{ \rule{3.2ex}{0.07em} }{ {} } }{ \Box\!\Box } $ & $(1, 1, 0)$ \\
         \hline\hline
    \end{tabular}
    \caption{Field content of the preon model. The fundamental fermions $\psi_I$, $I=Q,U,D,L,E,a$, with $a=1,2,3,4$, and the symmetric tensor fermion $\Omega$ are confined by $\SUp$ dynamics. 
    For realistic phenomenology, at least two $\SUF$ breaking spurions are required (if only one is included it can be diagonalized and the CKM matrix is trivial). 
    An example, which we will study in detail, is furnished by two scalars $\mathcal{A}$ and $\mathcal{A}'$ that mediate flavour physics through their Yukawa couplings 
    $\mathcal{A}^{(\prime)}\psi_a \psi_b$. 
    }
    \label{tab:model-def}
\end{table}
In this section, we review the essential features of the preon model, with a focus on the necessary ingredients needed for flavour physics. In particular, we will describe: {\it i)} the field content and how quarks and leptons emerge as composite bound states, {\it ii)} the role of the scalars $\mathcal{A}$ and $\mathcal{A}'$ in generating the flavour structure, and {\it iii)} the power counting rules for diagrams involving flavour violation.

\subsection{Field content} 
\label{sec:field-content}

The model is based on an $\SUp$ gauge theory that becomes strongly coupled at the ``preon scale'' $\Lambda_{\rm pre}$~\cite{Dobrescu:2021fny,Assi:2022jwg, Assi:2025rjx}. The matter content, summarized in Table~\ref{tab:model-def}, consists of chiral fermions transforming under $\SUp$ that also carry SM gauge charges, together with two scalar fields whose Yukawa couplings are responsible for breaking the $\SUF$ flavour symmetry.

The fermion content includes one left-handed Weyl fermion $\Omega$ transforming in the conjugate symmetric 2-tensor representation of $\SUp$ (dimension $\mathbf{120}$), and 19 left-handed Weyl fermions $\psi_I$ in the fundamental $\mathbf{15}$ representation. This field content 
is anomaly-free: the $\SUp$ gauge anomaly from $\Omega$ is cancelled by the 19 fundamentals, since the second Dynkin index is $T_2(\Omega) = 17/2$~\cite{Eichten:1982pn}. Furthermore, the model satisfies the $\text{U(1)}_\text{X}^3$ and $\text{U(1)}_{X} \times \SUp^2$ 't Hooft anomaly matching conditions\footnote{The gauge group $\text{U(1)}_\text{X}$ is a ``gedanken gauge field'' which can be made anomaly free via the addition of spectator fermions; it is not a dynamical field in the model.} that are a necessary consistency condition for the model to support the SM chiral fermions as bound states. 

Among the 19 fundamental fermions, five carry SM gauge charges corresponding to a single generation: $\psi_Q$ transforms as $(3, 2, +1/6)$, $\psi_U$ as $(\bar{3}, 1, -2/3)$, $\psi_D$ as $(\bar{3}, 1, +1/3)$, $\psi_L$ as $(1, 2, -1/2)$, and $\psi_E$ as $(1, 1, +1)$ under $\text{SU(3)}_\text{c} \times \text{SU(2)}_\text{L} \times \text{U(1)}_\text{Y}$. The remaining four preons, $\psi_a$ for $a=1,2,3,4$, are SM singlets whose interchange symmetry furnishes an $\SUF$ global flavour group. While confinement is predominantly driven by the strongly-coupled $\SUp$ dynamics, the SM gauge interactions act as a small perturbation that splits the bound state spectra (akin to the electromagnetic contribution to the $\pi^\pm, \pi^0$ mass difference in QCD). The resulting mass splitting is estimated using the most attractive channel 
heuristic \cite{Dimopoulos:1980hn,Dobrescu:2021fny}. 

To obtain the correct SM flavour structure, 
the model requires explicit breaking of the global $\SUF$ flavour group. The simplest implementation, adopted here, introduces two SM singlet scalar fields $\mathcal{A}$ and $\mathcal{A}'$ transforming in $\overline{\bf 105}$ and $\overline{\mathbf{120}}$ of $\SUp$, respectively, and 
whose masses are taken to be of order $\Lambda_{\rm pre}$.
Their Yukawa couplings to the singlet preons are given by
\begin{equation}
\mathcal{L}_{\rm Yuk} = \frac12\lambda_{ab} \mathcal{A}_{\alpha\beta} \psi_a^\alpha \psi_b^\beta + \frac12 \lambda'_{ab} \mathcal{A}'_{\alpha\beta} \psi_a^\alpha \psi_b^\beta + {\rm h.c.} \,,
\label{eq:Yukawa:A}
\end{equation}
where $a,b= 1,2,3,4,$ are the preon flavour indices labeling the four SM-singlet preons $\psi_a$, while $\alpha,\beta = 1,\ldots,15,$ are the $\SUp$ colour indices ($\mathcal{A}$ and $\mathcal{A}'$ carry two $\SUp$ colour indices).
The two $4\times 4$ complex matrices, the antisymmetric $\lambda_{ab}=-\lambda_{ba}$ and the symmetric $\lambda'_{ab}=\lambda'_{ba}$, are spurions that explicitly break the global $\SUF$ flavour group. By an $\SUF$ preon field redefinition the symmetric matrix $\lambda'$ can be brought to real-diagonal form via a Takagi decomposition; the antisymmetric matrix $\lambda$ then contains the residual physical flavour misalignment needed for a non-trivial CKM matrix. In practice neither spurion is taken in canonical form below: we instead impose the hierarchical Froggatt--Nielsen textures of \cref{eq:lambda_texture} on both, which is sufficient to capture the observed flavour structure.

Beyond the $\SUF$ flavour group, with $\psi_a$ forming a $\mathbf{4}$ of $\SUF$, the two scalar preon model also has another approximate global symmetry that has important consequences for the SM flavour structure. In the 
 limit where the SM hypercharge coupling is taken to zero, $\alpha_Y\rightarrow 0$,
 there is a global $\text{SU(2)}_\text{I}$ ``right-handed isospin'' symmetry under which $\psi_U$ and $\psi_D$ form a doublet.\footnote{The electroweak Higgs doublets $H_{ud}$, which we discuss in the next subsection, also form a doublet of $\text{SU(2)}_\text{I}$. Note furthermore, that $\lambda, \lambda'\ne 0$ do not break this symmetry; $\psi_a$ as well as ${\mathcal A}, {\mathcal A}'$ are singlets under $\text{SU(2)}_\text{I}$.} If this symmetry were exact, the up- and down-Yukawa matrices would be perfectly aligned and the CKM matrix trivial. In principle, there are two sources of misalignment: {\it i)} spontaneous breaking from the vevs of composite scalar fields, and {\it ii)} explicit breaking from hypercharge radiative corrections. We find that the first option more easily leads to a realistic CKM matrix, and we thus focus mainly on this possibility.

\subsection{Emergence of the three-generation Standard Model \label{sec:emergence-SM}}
The $\SUp$ gauge interactions confine at the scale $\Lambda_{\rm pre}$. Just as three quarks form a baryon in QCD, the confined spectrum here consists of three-preon bound states called prebaryons, of the form $\psi_I \psi_J \Omega$ and denoted as $\Omega_{IJ}$~\cite{Dobrescu:2021fny}. The existence of massless chiral prebaryons is supported by past large-$N$ analyses \cite{Eichten:1985fs},\!\footnote{The inclusion of scalars in the large representations, $\overline{\mathbf{105}}$ and $\overline{\mathbf{120}}$, may change some details in the large-$N$ limit.} and by the passing of non-trivial consistency conditions stemming from 't Hooft anomaly matching~\cite{tHooft:1979rat}. The SM quantum numbers of the massless fermions are inherited from the constituent preons. The key point for flavour physics is that each SM-charged preon can pair with any of the four singlets $\psi_a$ ($a=1,2,3,4$), so that the prebaryon spectrum initially contains four copies of each SM fermion species, as listed in Table~\ref{tab:prebaryons}.

The reduction from four to three generations is driven by the vev of a composite scalar $\phi_{1/6,a}$, which is a bound state of $\Omega_{DL}$ and $\Omega_{Qa}$ prebaryons, $\phi_{1/6,a}\sim (\Omega_{DL}\, \Omega_{Qa})$. The explicit breaking of $\SUF$ by $\lambda, \lambda'$
selects the vev to be in the $a=4$ direction in the flavour space. This vev allows the fourth copy of each SM species to pair with a prebaryon carrying conjugate quantum numbers (e.g., $\Omega_{Q4}$ pairs with $\Omega_{DL}$), forming a massive Dirac fermion that acquires a vectorlike mass $m_Q=y_Q\langle\phi_Q\rangle$ set by the vev of the corresponding composite scalar, which can lie well below $\Lambda_{\rm pre}$, and decouples \cite{Dobrescu:2021fny}. After all such pairings, exactly three chiral generations remain in the infrared.

The masses of the vectorlike fermions are set by the vevs of composite scalar bound states (di-prebaryons). The lightest such state is a vectorlike lepton weak singlet, which can be as light as a few hundred GeV~\cite{Assi:2022jwg}; the strongest LHC constraint on the spectrum comes from the vectorlike lepton doublet, excluded below $1045$~GeV~\cite{CMS:2022nty}.

The vectorlike fermions also affect electroweak precision observables through their mixing with the SM fermions. The vectorlike quarks whose masses enter the proton decay bound in \cref{eq:proton_bound}, $\mathcal Q_{7/6}$ and $\mathcal Q_{33}$, share no prebaryon with $H_{u,d}$, so their mixing with the SM quarks is not induced at leading order and precision data do not constrain them. The largest mixing involves the up-type vectorlike quarks, since $H_u$ contains one of their constituents and couples them to $q_{L3}$ with $y_*\approx 1.5$--$2$~\cite{Dobrescu:2021fny,Assi:2022jwg}, as in top-seesaw models~\cite{Dobrescu:1997nm,Chivukula:1998wd}. The mixing angle $\theta\simeq y_* v_u/(\sqrt2\, m)$ gives a positive contribution to the $T$ parameter that stays below $0.1$ for $m\gtrsim 4$~TeV~\cite{Lavoura:1992np,Chen:2017hak}, satisfied by their mass of about 30~TeV in the spectrum of Ref.~\cite{Assi:2022jwg}. The other vectorlike fermions couple to the SM fermions through $H_d$, so their mixing is suppressed by $\cos\beta$.

\begin{table}[t]
    \centering
    \renewcommand{\arraystretch}{1.4}
    \begin{tabular}{ccc}
    \hline\hline
         Prebaryon & Preon content & $SU(3)_c \times SU(2)_W \times U(1)_Y$ \\
         \hline
         $\Omega_{Qi}$ & $\psi_Q \psi_i \Omega$ & $3 \times (3, 2, +1/6)$ \\
         $\Omega_{Ui}$ & $\psi_U \psi_i \Omega$ & $3 \times (\bar{3}, 1, -2/3)$ \\
         $\Omega_{Di}$ & $\psi_D \psi_i \Omega$ & $3 \times (\bar{3}, 1, +1/3)$ \\
         $\Omega_{Li}$ & $\psi_L \psi_i \Omega$ & $3 \times (1, 2, -1/2)$ \\
         $\Omega_{Ei}$ & $\psi_E \psi_i \Omega$ & $3 \times (1, 1, +1)$ \\
         \hline
         $\Omega_{DL}$ & $\psi_D \psi_L \Omega$ & $(\bar{3}, 2, -1/6)$ \\
         $\Omega_{UL}$ & $\psi_U \psi_L \Omega$ & $(\bar{3}, 2, -7/6)$ \\
         \multicolumn{3}{c}{$\vdots$ (additional vectorlike states)}\\
         \hline\hline
    \end{tabular}
    \caption{Selected chiral prebaryons that form the SM fermions. The index $i=1,2,3,$ labels the SM generations after one linear combination of $\Omega_{Qa}$, denoted as $\Omega_{Q4}$, pairs with $\Omega_{DL}$ to form a massive vectorlike state. Similarly, one linear combination each for $\Omega_{Ua}$, 
         $\Omega_{Da}$, $\Omega_{La}$, and $\Omega_{Ea}$ pairs with the corresponding additional prebaryon (not shown) to form heavy vectorlike fermions. Other prebaryons with exotic SM quantum numbers, not shown above,
    also exist in the spectrum, as detailed in Ref.~\cite{Assi:2022jwg}.}
    \label{tab:prebaryons}
\end{table}

\subsection{The Higgs sector \label{sec:higgs-sector} }
The model also contains the necessary ingredients for a composite Higgs. Di-prebaryon bound states with the quantum numbers of a Higgs doublet can be formed in three ways: an up-type Higgs $H_u^{ab}=(\Omega_{Ua}\,\Omega_{Qb})$, a down-type Higgs $H_d^{ab}=(\Omega_{Da}\,\Omega_{Qb})$, and a lepton-type Higgs $H_\ell^{ab} = (\Omega_{La}\,\Omega_{Eb})$, where $a,b$ are flavour indices running over $1,\ldots,4$. Each type therefore admits a $4\times 4$ matrix of candidates.

Following the original proposal~\cite{Dobrescu:2021fny}, we identify $H_u^{43}$ and $H_d^{43}$ as the agents of electroweak symmetry breaking. Here the index $4$ corresponds to the singlet $\psi_a$ direction aligned with the $\phi_{1/6}$ vev, and $3$ to the heaviest chiral generation. The alternative $H_{u,d}^{33}$ would correspond to a purely third-generation (``top-composite'') Higgs, but this is disfavoured by electroweak precision data. The choice $H_{u,d}^{44}$ involves the decoupled fourth-generation direction and is phenomenologically viable but leads to a different Yukawa structure that we do not explore here.

The $H_u^{43}$ and $H_d^{43}$ fields define a type-II two-Higgs-doublet model (2HDM) with vacuum expectation values 
\beq
\label{eq:vud}
\langle H_{u,d} \rangle = v_{u,d}/\sqrt{2}~,
\eeq
 parametrized as $v_u = v\sin\beta$ and $v_d = v\cos\beta$ with $v = 246$~GeV and $\tan\beta = v_u/v_d$ treated as a free parameter. All other composite Higgs states are assumed to have positive masses above the TeV scale, so that they do not participate in electroweak symmetry breaking and are consistent with LHC constraints. The electroweak scale $v \ll \Lambda_{\rm pre}$ is parametrically small compared to the compositeness scale; as in other composite Higgs scenarios, this hierarchy requires that the model exhibits a second-order phase transition and that its parameters have been fine-tuned~\cite{Dobrescu:2021fny}. Since $\Lambda_{\rm pre}$ is set by proton decay, the tuning is larger than in composite Higgs models with compositeness at a few TeV~\cite{Dobrescu:1997nm,Chivukula:1998wd}, though much milder than in the original top-condensation models~\cite{Bardeen:1989ds}. The binding in the Higgs channel has to lie within about $(v/\Lambda_{\rm pre})^2\sim 10^{-11}$ of its critical value. This is a single condition on the strong dynamics, imposed after all radiative corrections are included, and the existence of a critical point (but not its precise location) is radiatively stable as long as the transition remains second order~\cite{Bardeen:1993pj}.

\begin{figure}[t]
  \centering
  \includegraphics[width=\textwidth]{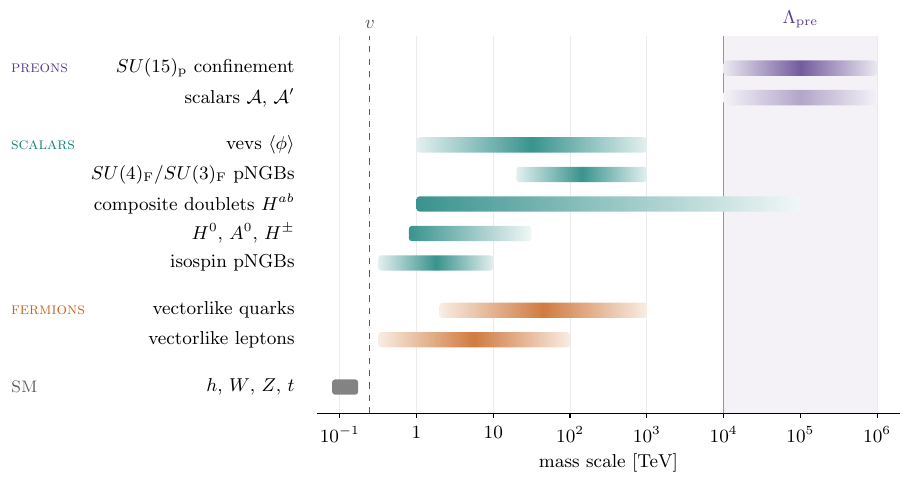}
  \caption{Schematic mass spectrum of the model. Bars show the expected mass ranges, and bars fading to the right have no definite upper value below $\Lambda_{\rm pre}$. The shaded band marks the range of lower bounds on $\Lambda_{\rm pre}$ from proton decay, and the dashed line the electroweak scale $v$. With the benchmark spurions, the flavour estimates favour the upper part of the vectorlike-fermion and pNGB ranges.}
  \label{fig:scales}
\end{figure}

 In the absence of fine tuning, all scalar masses are expected to lie close to $\Lambda_{\rm pre}$. When one state (the Higgs) has been tuned to be unnaturally light, we expect a hierarchical spectrum of scalar masses and vevs. The expectation, sometimes called Miransky scaling, is that the mass scale and vev of a state is $\sim \Lambda_{\rm pre} \exp[-C/\sqrt{g-g_*}]$ \cite{Miransky:1996pd,Braun:2010qs} where $C$ is some constant, $g$ is the effective binding strength, and $g_*$ is the critical binding at which that field's mass turns positive. The coupling $g$ depends on the strength of binding from SM gauge forces and the flavour-breaking Yukawas; it is therefore slightly different between different fields. One then expects a hierarchy of masses and vevs that are exponentially sensitive to these differences. If the transition in the Higgs channel is of this conformal type~\cite{Miransky:1996pd}, the same scaling also reduces the tuning of the electroweak scale, since $v\sim\Lambda_{\rm pre}\exp[-C/\sqrt{g-g_*}]$ then requires only $g-g_*\sim C^2/\ln^2(\Lambda_{\rm pre}/v)$, at the level of several percent for $C\sim\pi$. Whether the $\SUp$ dynamics realize this scaling is an open nonperturbative question. Following Ref.~\cite{Dobrescu:2021fny}, we assume a second-order transition. 
 The corresponding assumed spectrum of the model and the various scales involved are summarized in \cref{fig:scales}.

\subsection{Power counting and large-$N$ suppression \label{sec:large-N}}

The strong dynamics of $\SUp$ confinement cannot currently be computed from first principles using lattice methods, since no lattice formulation of chiral gauge theories yet exists. Instead, we rely on the large-$N$ expansion and naive dimensional analysis (NDA) to estimate the structure of effective operators below the confinement scale $\Lambda_{\rm pre}$. With $N=15$ the expansion parameter $1/N \approx 0.07$ is numerically small, providing reasonable control over the hierarchy of contributions. The large-$N$ counting rules for this class of $SU(N)$ theories with $N+4$ fundamentals and one conjugate symmetric fermion are developed in Ref.~\cite{Eichten:1985fs} (see also \cite{Manohar:1998xv} for a pedagogical discussion in the context of QCD).

When matching from preon-level operators to prebaryon-level effective operators at scales below $\Lambda_{\rm pre}$, we parametrize the amplitudes as
\begin{equation}
\mathcal{M} \sim \frac{C}{\Lambda_{\rm pre}^D} \times \left(\frac{1}{N}\right)^k \,,
\end{equation}
where $D$ is fixed by dimensional analysis and $C$ is, apart from flavour spurion $\lambda, \lambda'$ insertions, an order-unity matching coefficient that encodes non-perturbative dynamical information from the confining theory. The precise values of $C$ await future lattice calculations; for the present analysis we treat them as free parameters whose size we estimate with NDA. The integer $k$ counts the number of preon line interchanges between different prebaryons. A preon line interchange occurs when a constituent preon must be transferred from one bound state to another, requiring a reconnection of $\SUp$ colour lines analogous to OZI-suppressed processes in QCD \cite{Manohar:1998xv}. Each such interchange costs a factor of $1/N \approx 1/15$, so operators requiring multiple interchanges are strongly suppressed. Prebaryon propagators carry no additional $1/N$ suppression, so the leading contributions to any process are those with the fewest preon interchanges. 

\section{Quark and charged lepton masses}
\label{sec:masses}
The two-scalar preon model gives rise in the IR to Type-II 2HDM quark and charged lepton Yukawa couplings,\footnote{Yukawa couplings between $H_u$ and the down-quarks, and $H_d$ and the up-quarks, require scalar vevs (in analogy to the lepton-Yukawas) to absorb the mis-matched preons.}
\beq
\label{eq:Yukawas}
 \mathcal{L} \supset {Y}^{u}_{ij} \, \bar Q_i u_{Rj} H_u + {Y}^{d}_{ij} \, \bar Q_i d_{Rj}H_d + {Y}^{\ell}_{ij} \,\bar L_i \ell_{Rj} H_d+{\rm h.c.}.
 \eeq
After electroweak symmetry breaking these give the quark and charged lepton mass matrices $M^{u} = Y^{u}\,v_{u}/\sqrt{2}$, $M^{d} = Y^{d}\,v_{d}/\sqrt{2}$, and $M^{\ell} = Y^{\ell}\,v_{d}/\sqrt{2}$, with the Higgs vevs defined in \cref{eq:vud}.

 The above structure follows from four-prebaryon and six-prebaryon operators, which are generated when the scalars $\mathcal{A}$ and $\mathcal{A}'$ are integrated out at the preon-confinement scale, see \cref{fig:quark_diagrams}. The four-prebaryon operators (top row in \cref{fig:quark_diagrams}) contribute only to quarks, while the six-prebaryon operators contribute to both quark and charged lepton Yukawas (middle and bottom rows in \cref{fig:quark_diagrams} and \cref{fig:lepton_diagram}, respectively). Chirality restricts the topology of scalar-exchange diagrams that can contribute. Because the composite Higgses $H_{u,d}$ contain $\psi_3$ and $\psi_4$ constituents, every diagram must route these preons through the Yukawa vertices while connecting them to the external quark prebaryons carrying flavour indices $i,j\in\{1,2,3\}$, leading to the above Type II 2HDM structure.
 
The hierarchical structure of the SM Yukawas follows from an assumed hierarchical structure of the $\lambda, \lambda'$ coupling matrices of ${\mathcal A}, {\mathcal A}'$ scalars, \cref{eq:Yukawa:A}, for which we assume the following Froggatt--Nielsen-style~\cite{Froggatt:1978nt} texture in terms of a small parameter $\kappa\sim$~$0.17$,
\begin{equation}
\label{eq:lambda_texture}
 \lambda \sim \begin{pmatrix}
  0 & \kappa^{4} & \kappa^{3} & \kappa
  \\
 \kappa^{4} & 0 & \kappa^{2} & 1
 \\
 \kappa^{3} & \kappa^{2} & 0 & 1
 \\
 \kappa & 1 & 1 & 0
 \end{pmatrix},
\qquad
 \lambda' \sim \begin{pmatrix}
  \kappa^{4} & \kappa^{4} & \kappa^{3} & \kappa^{2}
  \\
 \kappa^{4} & \kappa^{2} & \kappa^{2} & \kappa
 \\
 \kappa^{3} & \kappa^{2} & 1 & 1
 \\
  \kappa^{2} & \kappa & 1 & 1
 \end{pmatrix}.
 \eeq
 The powers of $\kappa$ parametrize the hierarchy the spurions must have to reproduce the observed masses and mixings with order-one coefficients, in the same way as the hierarchical Yukawa couplings are inputs of the SM. Such a pattern can arise from a horizontal symmetry acting on the flavour preons $\psi_a$ and broken by a flavon vev, as in the Froggatt--Nielsen mechanism~\cite{Froggatt:1978nt}, however we do not attempt any detailed model building to engineer the textures in \cref{eq:lambda_texture} and simply adopt them as inputs.

 While both $\lambda$ and $\lambda'$ break the $\SUF$ flavour symmetry, this is not enough to obtain a realistic pattern of the SM quark masses. For instance, even in the limit where right-handed isospin symmetry $\SUI$ mixing $\psi_{U,D}$ remains unbroken, the above two spurions do lead to hierarchical quark masses. However, in this limit one predicts 
 \beq
 \label{eq:quarkmass:ratios}
 m_u/m_d=m_c/m_s=m_t/m_b,
 \eeq
  while in reality, for measured quark masses, this is only correct up to factors of a few. Furthermore, the CKM matrix would be predicted to be $V_\text{CKM}=1$, in contradiction with the measurements. If the $\SUI$ is broken only by hypercharge SM interactions, the above quark mass relations would only get corrections of order ${\mathcal O}(\alpha_Y/\pi)\lesssim 10^{-2}$, which is not enough to bring \cref{eq:quarkmass:ratios} in line with observations, and similarly not for $V_\text{CKM}$ to agree with observations. 
 
 This leads us to conclude that the observed pattern of quark masses and mixings requires the $\SUI$ to also be broken by vevs of composite scalars, $\phi_{7/6}\sim (\Omega_{QE} \Omega_{UL})$, and $\phi_{Q}\sim (\Omega_{Q4} \Omega_{DL})$. The corresponding contributions to $Y^u$ and $Y^d$ Yukawas are shown in the middle and bottom rows in \cref{fig:quark_diagrams}. While the $\lambda, \lambda'$ spurion structure of the contributions to $Y^u$ and $Y^d$ Yukawas are the same in the two cases, they are proportional to different overall prefactors, $\propto \langle \phi_{7/6}\rangle^2$, and $\propto\langle \phi_{Q}\rangle^2$, breaking the $\SUI$ symmetry.\footnote{ More precisely, the contributions are proportional to $y_{\phi_i} \langle \phi_i \rangle=m_{Q_i}$, where $y_{\phi_i}$ is the Yukawa coupling between $\phi_i$ and the vector-like fermions $Q_i$, constituents of $\phi_i$ \cite{Assi:2025rjx}. } Similarly, the vevs of $\phi_{7/6}$ and $\phi_{88}\sim \Omega'_{QU} \Omega'_{QD}$ give rise to a realistic pattern of charged lepton masses. In the remainder of this section we give further details on the quark and charged lepton mass generation and their parametric scalings.

 \subsection{The quark sector}
 \label{sec:quark:masses}

\begin{figure}[t!]
  \begin{center}
\includegraphics[width=0.45\textwidth]{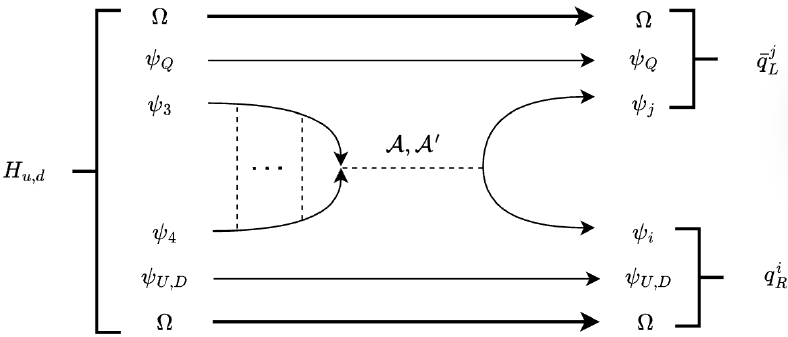}
\includegraphics[width=0.45\textwidth]{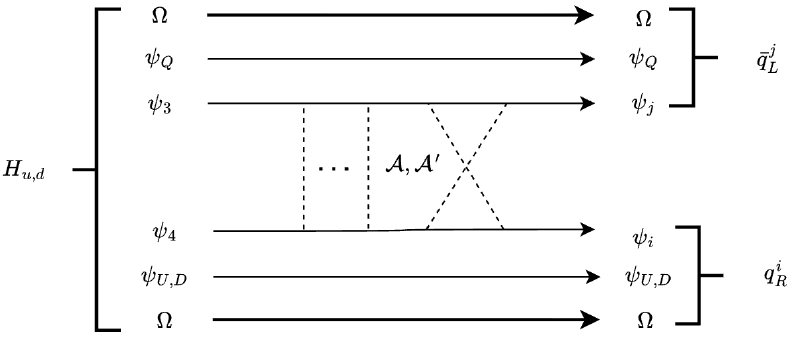}
  \\
  \includegraphics[width=0.90\textwidth]{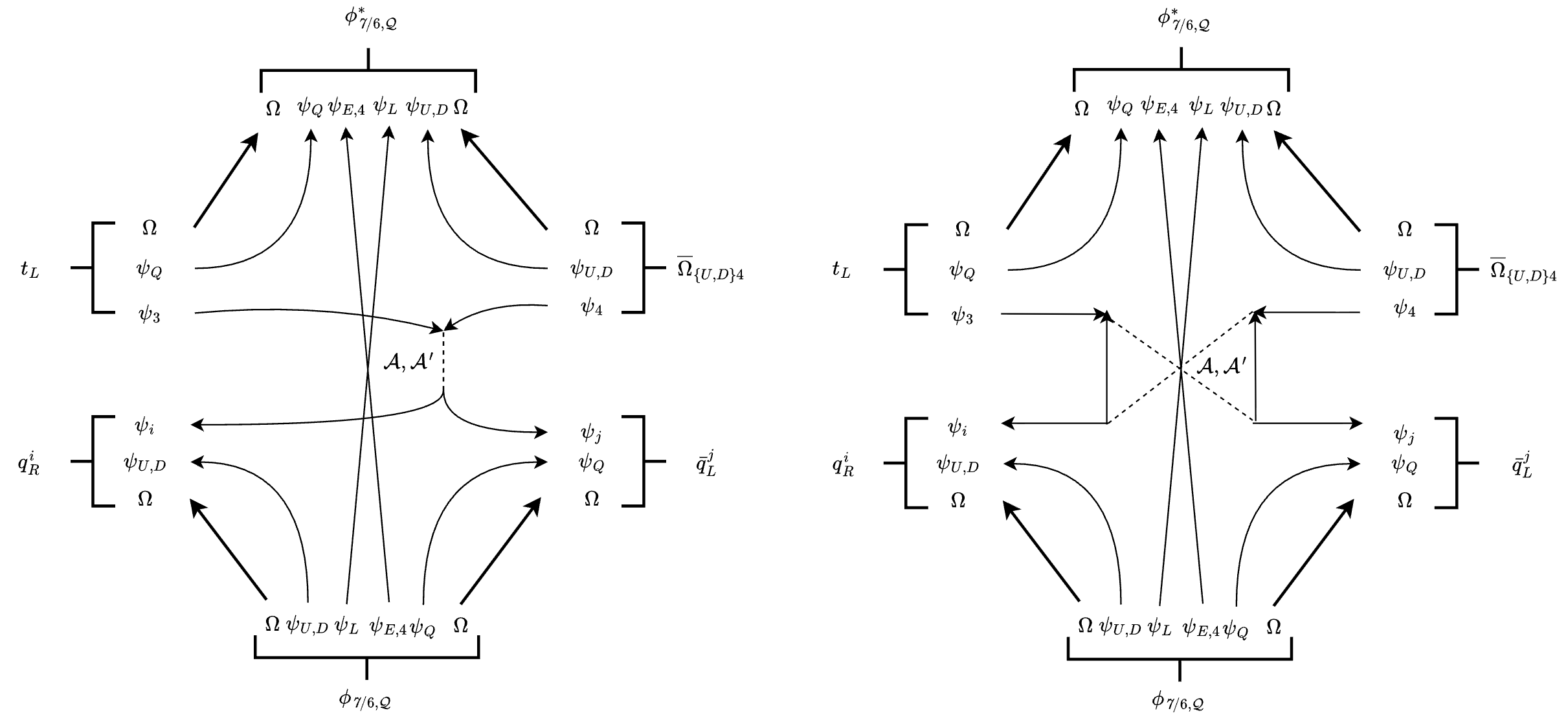} 
\caption{Diagram topologies contributing to the quark Yukawa couplings in the
perturbative $\lambda,\lambda'$ regime.
{\bf Top:} the $s$-channel tree diagram (left) with a single $\mathcal{A}$ or
$\mathcal{A}'$ exchange between the composite Higgs and the external quark
prebaryons, and one loop corrections with $\mathcal{A}$ or
$\mathcal{A}'$ exchanged in a $t$ channel (right). 
{\bf Bottom:} the corrections to the up- or down-quark Yukawas from vevs of $\phi_{7/6}$ and $\phi_Q$ that come from diagrams with either single or double $\mathcal{A}, \mathcal{A}'$ insertions (note that $t_L$ and $\bar \Omega_{\{U,D\}4}$ form $H_{u,d}$, respectively).
}
  \label{fig:quark_diagrams}
  \end{center}
\end{figure}

The key challenge is to extract quantitative predictions from the strongly-coupled $\SUp$ dynamics. As discussed above, we allow the non-perturbative matrix elements to vary and search for benchmark solutions that reproduce the observed fermion spectrum. The $\lambda, \lambda'$ couplings are treated as perturbative, and we keep only the leading contributions in the spurion insertions. With these caveats, the SM Yukawas for up and down quarks, \cref{eq:Yukawas}, are given by
\begin{equation}
    \begin{split}
    \label{eq:quark_yukawa}
    Y^{u,d}_{ij} =& F_{u,d}' 
        \lambda'_{43}\lambda_{ij}^{\prime *} +
      F_{u,d} 
        \lambda_{43}\lambda_{ij}^{*}
    + \frac{1}{16\pi^2} \Big[ G_{u,d} \lambda_{3a} \lambda_{ia}^* \lambda_{4b}\lambda_{jb}^*+ I_{u,d} \lambda'_{3a} \lambda_{ia}^* \lambda'_{4b}\lambda_{jb}^*
    \\
    &\qquad\qquad\qquad\qquad\qquad\qquad\quad + J_{u,d} \lambda_{3a} \lambda_{ia}^{\prime *} \lambda_{4b}\lambda_{jb}^{\prime*}+ K_{u,d} \lambda'_{3a} \lambda_{ia}^{\prime*} \lambda'_{4b}\lambda_{jb}^{\prime *}\Big],
    \end{split}
\end{equation}
where the summation over repeated indices $a,b,=1,\ldots,4$ is understood, while, as usual, $i, j=1,2,3$. 

The real-valued nonperturbative ``hadronic'' matrix elements $F_{u,d}, F_{u,d}'$ in \cref{eq:quark_yukawa} encode the nonperturbative $\SUp$ dynamics for contributions with a single ${\mathcal A}$ or ${\mathcal A}'$ exchange in the $s-$channel, respectively, shown in the top left panel in \cref{fig:quark_diagrams}. Since ${\mathcal A}$ and ${\mathcal A}'$ are in different representations of $\SUp$, we expect $F_{u,d}\ne F_{u,d}'$. 
Furthermore, if only the diagram in the top left panel in \cref{fig:quark_diagrams} contributes, then $\SUI$ requires $F_u=F_d$ and $F_u'=F_d'$, up to small electroweak corrections. However, if the composite scalars, $\phi_{7/6}$ and $\phi_Q$, obtain vevs, then also the diagrams in middle left and bottom left panels in \cref{fig:quark_diagrams} are relevant. 
The corresponding effective operators, matched at $\Lambda_{\rm pre}$, take the form
\begin{equation}
\label{eq:quark:vev:operator}
\frac{|\langle \phi_{7/6} \rangle|^2}{N \Lambda_{\rm pre}^2} \underbrace{\left(\bar{\Omega}_{Qj} \bar{\Omega}_{Ui}\right)}_{\bar q_L^j u_R^i} \underbrace{\left( \Omega_{Q3} \Omega_{U4}\right)}_{H_u} \times \lambda_{34}^{(\prime)}\lambda_{ij}^{(\prime)*}, \qquad \frac{|\langle \phi_{Q} \rangle|^2}{N \Lambda_{\rm pre}^2} \underbrace{\left(\bar{\Omega}_{Qj} \bar{\Omega}_{Di}\right)}_{\bar q_L^j d_R^i} \underbrace{\left( \Omega_{Q3} \Omega_{D4}\right)}_{H_d} \times \lambda_{34}^{(\prime)}\lambda_{ij}^{(\prime)*}~,
\end{equation}
with $\lambda$ ($\lambda'$) spurion insertions arising in the case of ${\mathcal A}$ (${\mathcal A}'$) exchanges. 

The contributions are $1/N$ suppressed, but also contains vevs of light bound states, $\phi_{7/6}$ and $\phi_Q$, the di-prebaryon scalars that are part of the physical spectrum below the confinement scale.
Their condensation gives masses $m_{Q_i}=y_{\phi_i}\langle \phi_i\rangle$ to the accompanying composite vectorlike quarks, which can lie well below $\Lambda_{\rm pre}$ (cf.~\cref{sec:proton}). The $1/N$ suppression can then be compensated either by vevs somewhat above $\Lambda_{\rm pre}$, or by non-local contributions in which these light states propagate below the matching scale with either effect absorbed into the nonperturbative coefficients.
The presence of such light bound states below $\Lambda_{\rm pre}$ is a recurring feature of the construction: we will encounter it again for the lepton Yukawas (\cref{sec:lepton:sector}), for the pNGBs of $\SUF$ breaking (\cref{sec:pNGB}), and in the proton-decay bounds of \cref{sec:flavour}, which depend explicitly on the vectorlike-quark masses.
 These contributions are phenomenologically very important: they are proportional to the same two spurion insertions, but misalign up and down quark Yukawas. Once absorbed into the nonperturbative factors they lead to $F_u\ne F_d$ and $F_u'\ne F_d'$, where we parametrize the differences as
\beq
F_u=F_d(1+\delta_F), \qquad {\rm and} \qquad F_u'=F_d'(1+\delta_{F'})~. 
\eeq

While $F_{u,d}'$ and $F_{u,d}$ contributions suffice to obtain the parametric size of up- and down- quark matrices, as well as the CKM matrix, in order to match the actual numerical values we also need to include the contributions that arise at one loop order, with four insertions of $\lambda, \lambda'$ spurions (right column panels in \cref{fig:quark_diagrams}). These contributions are down by $N/16\pi^2$ factor and are parametrized by the real nonperturbative matrix elements $G_{u,d}$, $I_{u,d}$, $J_{u,d}$, and $K_{u,d}$. 
Since ${\mathcal A}$ and ${\mathcal A}'$ are in different representations of $\SUp$, we expect $G_{u,d}\ne I_{u,d}\ne J_{u,d} \ne K_{u,d}$, while the contributions proportional to the composite scalar vevs (middle right and bottom right panels in \cref{fig:quark_diagrams}) ensure $G_u\ne G_d$, $I_u\ne I_d$, $J_u\ne J_d$, $K_u\ne K_d$ (in complete analogy with the tree level exchanges). We parametrize the differences as
\beq
G_u=G_d(1+\delta_G), \quad I_u=I_d(1+\delta_{I}), \quad J_u=J_d(1+\delta_{J}), \quad K_u=K_d(1+\delta_{K}).
\eeq

The physical quark masses and CKM matrix are obtained by diagonalizing the Yukawa matrices via singular value decomposition (SVD),
\begin{equation}
\label{eq:Yud:rotate}
Y^u = U_u \, {\rm diag}(y_1^u, y_2^u, y_3^u) \, V_u^\dagger, \quad
Y^d = U_d \, {\rm diag}(y_1^d, y_2^d, y_3^d) \, V_d^\dagger,
\end{equation}
giving the physical quark masses 
\begin{equation}
\label{eq:physical:quark:masses}
    m_{u,c,t} = \frac{v_u}{\sqrt 2} \, y_{1,2,3}^u, \quad
    m_{d,s,b} = \frac{v_d}{\sqrt 2} \, y_{1,2,3}^d.
\end{equation}
The CKM matrix arises from the misalignment of left-handed rotations,
\begin{equation}
    V_{\rm CKM} = U_u^\dagger U_d,
\end{equation}
where the complex coefficients in $\lambda, \lambda'$ combine into a single physical 
 { CP}-violating phase,
 measured by the Jarlskog invariant
\begin{equation}
    \mathcal{J} = {\rm Im}\left[V_{us} V_{cb} V_{ub}^* V_{cs}^*\right].
\end{equation}

Treating all nonperturbative matrix elements in \cref{eq:quark_yukawa} as $F_{u,d}^{(\prime)},G_{u,d}, I_{u,d}$, $J_{u,d}, K_{u,d}\sim {\mathcal O}(1)$, and the loop factor as $N/(16\pi^2)\sim {\mathcal O}(\kappa^2)$, this gives a parametric prediction for the up and down quark Yukawas and for the CKM matrix,\footnote{Numerically, $N/(16\pi^2)=0.09$ is somewhere between ${\mathcal O}(\kappa^2)$ and ${\mathcal O}(\kappa)$. In our benchmark, $F_{u,d}^{(\prime)}$ are somewhat larger than the nonperturbative parameters entering the one loop corrections, which can thus be treated as though they are parametrically ${\mathcal O}(\kappa^2)$ suppressed. Alternatively, if one treats the loop factor as $N/(16\pi^2)={\mathcal O}(\kappa)$, this gives
\begin{equation}
\label{eq:hatVCKM:alter}
\hat Y_u\sim \hat Y_d \sim \begin{pmatrix}
  \kappa^{3} & \kappa^{2} & \kappa^{2}
  \\
 \kappa^{2} & \kappa & \kappa
 \\
 \kappa^{2} & \kappa & 1
 \end{pmatrix},
 \qquad
\hat V_\text{CKM} \sim
 \begin{pmatrix}
  1 & \kappa & \kappa^{2}
  \\
 \kappa & 1 & \kappa^{}
 \\
 \kappa^{2} & \kappa^{} & 1
 \end{pmatrix}.
\end{equation}
This signals that some amount of cancellation, on the order of factor of few, between different terms is required, in order to obtain the observed CKM matrix. 
}
\begin{equation}
\label{eq:hatVCKM}
\hat Y_u\sim \hat Y_d \sim \begin{pmatrix}
  \kappa^{4} & \kappa^{3} & \kappa^{3}
  \\
 \kappa^{3} & \kappa^{2} & \kappa^{2}
 \\
 \kappa^{3} & \kappa^{2} & 1
 \end{pmatrix},
 \qquad
\hat V_\text{CKM} \sim
 \begin{pmatrix}
  1 & \kappa & \kappa^{3}
  \\
 \kappa & 1 & \kappa^{2}
 \\
 \kappa^{3} & \kappa^{2} & 1
 \end{pmatrix}.
\end{equation}
Here, the one-loop contributions enhance the $[\hat Y_{u,d}]_{12}\sim [\hat Y_{u,d}]_{21}$ Yukawa matrix elements from ${\mathcal O}(\kappa^4)$ size at tree level, to ${\mathcal O}(\kappa^3)$ after one-loop contributions. This also makes $V_{us,cd}$ larger, ${\mathcal O}(\kappa)$ as in the observed CKM, instead of ${\mathcal O}(\kappa^2)$ that would be expected from only tree level contributions. The mixing matrices that diagonalize the Yukawas, $U_{u,d}, V_{u,d}$ are expected to have a hierarchical structure parametrically similar to the CKM matrix, 
\beq
\label{eq:mixing:matrices:parametric}
\hat U_{u,d} \sim \hat V_{u,d} \sim \hat V_\text{CKM},
\eeq
with $\hat V_\text{CKM}$ given in \cref{eq:hatVCKM}.

The diagonal SM Yukawas are predicted to be
\beq
\hat y_1^u, \hat y_1^d\sim \kappa^{4}, \quad \hat y_2^u, \hat y_2^d\sim \kappa^{2}, \quad \hat y_3^u, \hat y_3^d\sim 1.
\eeq
Assuming that $v_u/v_d=\tan\beta\sim 1/\kappa^{2}$, as will be the case for our benchmark, this then gives for the parametric sizes of the predicted quark masses (in terms of $v/\sqrt 2$)
\beq
\label{eq:mass:scalings:naive}
\hat m_u\sim \kappa^4, \, \hat m_c\sim \kappa^2, \, \hat m_t\sim 1, \qquad \hat m_d\sim \kappa^6, \,\hat m_s\sim \kappa^4, \,\hat m_b\sim \kappa^2. 
\eeq
These parametric estimates describe quite well the experimental situation, except for $m_u$ and $m_c$, which are numerically closer to 
\beq
m_u\sim \kappa^7, \quad m_c\sim \kappa^3.
\eeq

In the numerical benchmark that we will use to estimate flavour constraints on $\Lambda_\text{pre}$, the above parametric scaling is reasonably well reproduced, up to several cancellations of order few between various non-perturbative parameters (see App. \ref{app:parameters} for details). 
Note that the difference between top and bottom quark masses is explained predominantly by $v_u\gg v_d$, cf. \cref{eq:physical:quark:masses}, with the remainder due to the difference between $F_u\ne F_d$. The scaling in \cref{eq:mass:scalings:naive} predicts $m_c/m_s\sim v_u/v_d\gg 1$, which agrees reasonably with experiment, but also $m_u/m_d\sim v_u/v_d\gg 1$, which does not. To match the observed values, $m_u\lesssim m_d$, thus requires cancellations between different nonperturbative contributions to the level of about a few$\times10^{-2}$ (in contrast to $m_c$ that only requires a cancellation on the order of few). 

In \cref{app:parameters} we give the benchmark values for the nonperturbative parameters and values of $\lambda, \lambda'$ matrix elements that reproduce the SM quark masses and CKM matrix. We also give the benchmark numerical values of diagonalization matrices $\hat U_{u,d} \sim \hat V_{u,d}$. These have a parametric form as in \cref{eq:mixing:matrices:parametric}, up to a few numerical differences that are above a factor of a few, see \cref{app:parameters}. The sensitivity of those results to small perturbations of the fitted inputs is quantified in \cref{app:fine-tuning}.

\subsection{The lepton sector}
\label{sec:lepton:sector}

\begin{figure}[t!]
  \begin{center}
\includegraphics[width=0.95\textwidth]{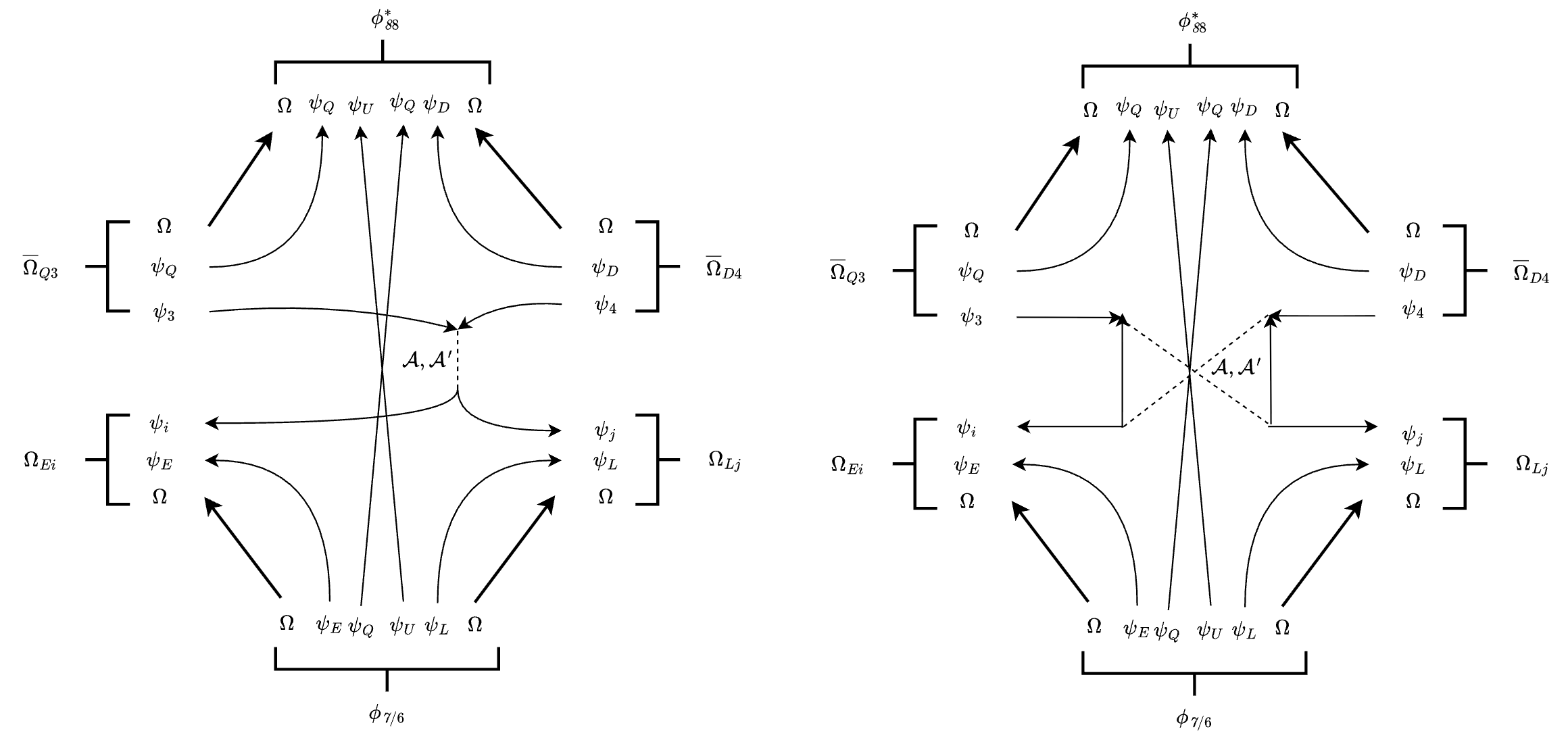}   
   \caption{One of the diagram topologies contributing to charged lepton Yukawa couplings. The effective operator connects the composite Higgs doublet $H_d$ to lepton prebaryons through di-prebaryon vevs $\langle\phi_{7/6}\rangle$ (bound state of $\Omega_{QE}$ and $\Omega_{UL}$) and $\langle\phi_{88}\rangle$ (bound state of $\Omega'_{QU}$ and $\Omega'_{QD}$), with a single $\mathcal{A}$ or $\mathcal{A}'$ exchange providing the leading $\lambda'_{34}\lambda'^{*}_{ij}$ flavour structure. Additional topologies with two $\mathcal{A}^{(\prime)}$ exchanges generate the four box channels $G^\ell, I^\ell, J^\ell, K^\ell$ in \cref{eq:lepton_yukawa}, weighted by the loop factor $N/(16\pi^{2})$. }
  \label{fig:lepton_diagram}
  \end{center}
\end{figure}

The charged lepton Yukawa couplings differ qualitatively from the quark sector. Because the lepton prebaryons $\Omega_L$ and $\Omega_E$ do not share a common preon species with the composite Higgs $H_d \sim (\bar\Omega_{D4}\Omega_{Q3})$, there is no ${\mathcal O}(N^0)$ diagram such as the one in the top left panel in \cref{fig:quark_diagrams} for quarks. 
Lepton Yukawas are instead necessarily proportional to the vevs of composite scalars, and are $1/N$ suppressed. 

If the only contribution to lepton masses were a local operator generated at the scale $\Lambda_{\rm pre}$ the tau-lepton mass would be too small. Therefore, as anticipated in \cref{sec:quark:masses}, light bound states of the $\SUp$ dynamics are required to play an essential role here. In particular, the ``non-local'' contributions mediated by the composite doublets $H^\ell_{ij}\sim (\Omega_{Li} \Omega_{Ej})$, whose masses can lie parametrically below $\Lambda_{\rm pre}$, are required. This can be understood via a two-step matching procedure. 

First, consider matching at the scale $\Lambda_{\rm pre}$. The contributing diagram topologies are shown in Figure~\ref{fig:lepton_diagram}. The diagram in the left panel in \cref{fig:lepton_diagram}, matched at $\Lambda_{\rm pre}$, results in an effective operator of the form
\begin{equation}
\label{eq:lepton:mass:operator}
\frac{\langle \phi_{7/6} \rangle \langle \phi_{88} \rangle}{N \Lambda_{\rm pre}^2} \underbrace{\left(\Omega_{D4} \Omega_{Q3}\right)}_{H_d} \underbrace{\left(\bar{\Omega}_{Li} \bar{\Omega}_{Ej}\right)}_{\bar L_i \ell_{Rj}} \times \lambda'_{34}\lambda_{ij}^{\prime\textcolor{black}{*}},
\end{equation}
in complete analogy with the ${\mathcal O}(1/N)$ contributions to the quark Yukawas in \cref{eq:quark:vev:operator}.
The composite scalars $\phi_{7/6}$ and $\phi_{88}$ are di-prebaryon bound states $(\Omega_{QE} \Omega_{UL})$ and $(\Omega'_{QU} \Omega'_{QD})$, respectively, the bilinear $({\Omega}_{D4} \Omega_{Q3})$ forms the composite Higgs $H_d$, while $({\Omega}_{Li} \Omega_{Ej})$ carries the lepton flavour indices, and results in the $\bar L_i \ell_{Rj}$ scalar current. 
The above matching calculation fixes the quartic $H^\ell H_d \phi_{88} \phi_{7/6}$ coupling.

Next, consider \cref{fig:lepton_nonlocal}, i.e., the non-local contribution in the theory where $H^\ell$ is a propagating degree of freedom. With $\phi_{88}$ and $\phi_{7/6}$ getting vevs, there is effectively mixing between $H^\ell$ and $H^d$. Integrating out $H^\ell$ then generates the same lepton-mass operator as in \cref{eq:lepton:mass:operator}, but with the suppression scale replaced by $\Lambda_{\rm pre} \rightarrow M_{H^\ell}$. This topology is illustrated in \cref{fig:lepton_nonlocal}. Its contributions are absorbed in the unknown nonperturbative functions $F_\ell, F_\ell'$. 

The one-loop suppressed contributions to the quartic $H^\ell H^d \phi_{88}\phi_{7/6}$ coupling, shown in right panel in \cref{fig:lepton_diagram}, involve four insertions of $\lambda, \lambda'$ spurions. As in the quark case, \cref{eq:quark_yukawa}, we introduce nonperturbative matrix elements for separate contributions from the various diagrams in \cref{fig:lepton_diagram}, keeping the spurion insertions 
and loop factor scalings explicit. 
The lepton Yukawa matrix is thus given by
\begin{equation}
    \begin{split}
    \label{eq:lepton_yukawa}
    Y^{\ell}_{ij} =& F_{\ell}' 
        \lambda'_{43}\lambda_{ij}^{\prime *} +
      F_{\ell} 
        \lambda_{43}\lambda_{ij}^{*}
    + \frac{1}{16\pi^2} \Big[ G_{\ell} \lambda_{3a} \lambda_{ia}^* \lambda_{4b}\lambda_{jb}^*+ I_{\ell} \lambda'_{3a} \lambda_{ia}^* \lambda'_{4b}\lambda_{jb}^*
    \\
    &\qquad\qquad\qquad\qquad\qquad\qquad\quad + J_{\ell} \lambda_{3a} \lambda_{ia}^{\prime *} \lambda_{4b}\lambda_{jb}^{\prime*}+ K_{\ell} \lambda'_{3a} \lambda_{ia}^{\prime*} \lambda'_{4b}\lambda_{jb}^{\prime *}\Big],
    \end{split}
\end{equation}
where the summation over repeated indices $a,b,=1,\ldots,4$ is understood. Since the nonperturbative functions include dynamics of different prebaryon bound states than for quark Yukawas, they differ from the quark ones, i.e., $F_\ell\ne F_{u,d}$, etc. In our numerics, when determining the benchmark values of $\lambda, \lambda'$ and the nonperturbative functions, we assume for simplicity that these are of similar size for quarks and leptons, i.e., $F_\ell \sim F_u \sim F_d$, etc. The result is a phenomenologically viable benchmark, though one could relax further the assumptions entering our numerics in future studies. 

\begin{figure}[t!]
  \begin{center}
  \includegraphics[width=0.75\textwidth]{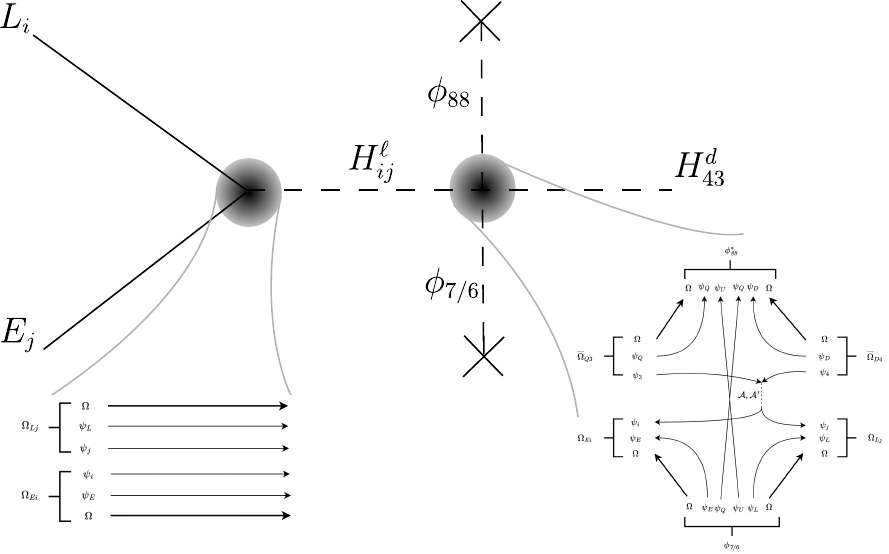}
  \caption{Non-local contribution to the charged lepton Yukawa. The lepton prebaryons $\Omega_{Li}$ and $\Omega_{Ej}$ form the composite scalar $H^\ell_{ij}\sim(\bar\Omega_{Li}\Omega_{Ej})$, which propagates with mass $M_{H^\ell}$ and mixes into the composite down-type Higgs $H_d\sim(\bar\Omega_{D4}\Omega_{Q3})$ through the quartic $\phi_{88}\phi_{7/6}H^dH^\ell$ (filled dot), with the two scalars $\phi_{7/6}$ and $\phi_{88}$ replaced by their vevs (crosses). Integrating out $H^\ell$ generates the same lepton-mass operator as the local diagram in \cref{eq:lepton:mass:operator}, but with the suppression scale $\Lambda_{\rm pre}$ replaced by $M_{H^\ell}$; when $M_{H^\ell}\ll \Lambda_{\rm pre}$ this non-local channel dominates over the local one and can provide an $\mathcal O(1)$ effective coupling without requiring $\phi$-vevs near $\Lambda_{\rm pre}$. The $\lambda'$ spurion insertions required by $\SUF$ breaking ($\lambda'_{34}$ and $\lambda'^*_{ij}$) are implicit in the composite-Higgs formation at each end and are not drawn.}
  \label{fig:lepton_nonlocal}
  \end{center}
\end{figure}

The charged lepton masses follow from the SVD diagonalization of $Y_\ell$, 
\begin{equation}
\label{eq:Yell:rotate}
Y_\ell = U_\ell \, {\rm diag}(y_1^\ell, y_2^\ell, y_3^\ell) \, V_\ell^\dagger, \quad
m_{e,\mu,\tau} = \frac{v_d}{\sqrt 2} \, y_{1,2,3}^\ell,
\end{equation}
where we order the singular values in increasing mass. Parametrically, we expect the charged lepton matrix to be similar to $Y_{u,d}$ Yukawa matrices, 
\cref{eq:hatVCKM},
\beq
\label{eq:lepton:Yukawa:scaling}
\hat Y_\ell\sim \begin{pmatrix}
  \kappa^{4} & \kappa^{3} & \kappa^{3}
  \\
 \kappa^{3} & \kappa^{2} & \kappa^{2}
 \\
 \kappa^{3} & \kappa^{2} & 1
 \end{pmatrix},
\eeq
so that after diagonalization by 
\beq
\label{eq:mixing:matrices:parametric:leptons}
\hat U_{\ell} \sim \hat V_{\ell} \sim \hat V_\text{CKM},
\eeq
the diagonal lepton Yukawa matrices and masses are expected to be 
\beq
\hat y_1^\ell \sim \kappa^{4}, \, \hat y_2^\ell \sim \kappa^{2}, \, \hat y_3^\ell \sim 1, \quad\text{and}\quad \hat m_e \sim \kappa^{6}, \, \hat m_\mu \sim \kappa^{4}, \, \hat m_\tau \sim \kappa^2,
\eeq
assuming $\tan\beta\sim 1/\kappa^2$. This agrees quite well with the experimental values for charged lepton masses, except for electron mass, for which some additional suppression from several nonperturbative contributions is needed to give the experimentally observed value, which is closer to $m_e\sim \kappa^7$.

The unitary matrices $U_\ell, V_\ell$ rotate from the interaction basis to the lepton mass basis, transforming $\lambda, \lambda'$ spurions into several distinct matrices that 
then enter the flavour constraints on $\Lambda_\text{pre}$ dynamics, which we discuss in more detail in \cref{sec:flavour}. The benchmark values for $U_\ell, V_\ell$ are hierarchical, but differ from the parametric expectation in \cref{eq:mixing:matrices:parametric:leptons} by up to an order of magnitude, for details see App. \ref{app:parameters}. For this reason we will give both the naive parametric expectations for FCNCs involving leptons, as well as the expectations based on the benchmark values of mixing angles. Note that we do not attempt to model the PMNS matrix which we expect to come from a type-I see-saw with the heavy neutral leptons $\psi_i \Omega \psi_j$.

\subsection{Aside: the fate of Nambu-Goldstone bosons}
\label{sec:pNGB}

The spontaneous breaking $\SUF\rightarrow SU(3)_{\rm F}$ by the vev of $\phi_{1/6}$ leaves eight generators unbroken. Then, since $\dim SU(4)-\dim SU(3)=15-8=7$, the breaking produces seven massless Goldstone bosons in the $\SUF$ symmetric limit. However, $\SUF$ is explicitly broken by the Yukawa couplings $\lambda$ and $\lambda'$, so the would-be Goldstones become pseudo Nambu-Goldstone bosons (pNGBs) with masses set by the size of the explicit breaking.

Under the unbroken $SU(3)_F$, the seven pNGBs decompose as $\bar{\bf 3}\oplus {\bf 3} \oplus {\bf 1}$. The dominant source of explicit breaking is $\lambda'_{44}$, which transforms as a singlet under $SU(3)_F$. Because a singlet couples to all $SU(3)_F$ representations, it contributes to the masses of all seven pNGBs. In the current benchmark, the large top Yukawa is generated by the 3--4 spurion block, leading to $\lambda'_{44}\sim O(1)$, and therefore the pNGBs are not parametrically light. Instead, their masses are of order $\langle \phi_{1/6} \rangle$, well above the electroweak scale but potentially below $\Lambda_{\rm pre}$ (this may impact some of the proton decay modes proposed in \cite{Assi:2022jwg}). In particular, unless $\langle\phi_{1/6}\rangle$ lies far below $\Lambda_{\rm pre}$, they are too heavy to mediate long-distance flavour-violating processes that would compete with the local operators analysed in Section~\ref{sec:flavour}.

The lightest pNGBs in the spectrum are those associated to the spontaneously broken $U\leftrightarrow D$ isospin symmetry. The only source of explicit breaking here comes from hypercharge assignments, and the masses will therefore be small $m^2 \sim \Delta (\alpha_Y/4\pi) \langle \phi_{1/6} \rangle^2 \ll \langle \phi_{1/6} \rangle^2$, where $\Delta=1/6$ if dominated by hypercharge exchange with $\psi_Q$ or $\Delta=4/9 - 1/9=1/3$ if dominated by self-energy like topologies. These pNGBs may then plausibly lie two orders of magnitude $\big($taken from $\sqrt{(1/6) \times \alpha/(4\pi)}$ $\big)$ below the scale of $\langle \phi_{1/6}\rangle$ and therefore represent collider targets if $\langle \phi_{1/6}\rangle \lesssim 100~{\rm TeV}$.

\section{Phenomenology of low energy constraints} 
\label{sec:flavour}
The main constraints on the preon model are expected to come from indirect constraints. Ref.~\cite{Assi:2022jwg} found that the proton decay limits imply $\Lambda_\text{pre}\gtrsim 10^4$\,TeV. This is well above direct collider reach (i.e., where quark and lepton substructure would reveal itself). Such high scales can, however, be probed indirectly by various flavour transitions. After reviewing the proton decay constraints, we then estimate the sensitivity to $\Lambda_\text{pre}$ for a number of different flavour changing neutral current (FCNC) transitions.

The new physics FCNCs are generated by $\mathcal{A}$ and $\mathcal{A}'$ exchanges, so that the flavour physics phenomenology of the model is determined by the couplings $\lambda'$ and $\lambda$, evaluated in one of the three natural low-energy bases (up-quark, down-quark, or charged lepton mass bases). For the numerical benchmark these are given in Appendix \ref{app:parameters}.

\subsection{Proton decay \label{sec:proton}}

The $\SUp$ dynamics generates an operator with eight prebaryon legs, four of which get combined into two composite scalars, $\phi_{7/6}\sim (\Omega_{UL}\Omega_{QE})$, and $\phi_{33}\sim (\Omega_{QQ} \Omega_{LQ})$~\cite{Assi:2022jwg}. 
  The diagram is nonplanar, suppressed by $1/(N^3 N_c^2)$, and reduces to a baryon number violating dimension-six operator $(u_L d_L)(u_Re_R)$,
  once the composite scalars $\phi_{7/6}$ and $\phi_{33}$ are replaced by their vevs (the vevs violate both baryon and lepton number).\footnote{Reference~\cite{Assi:2022jwg} also identifies a possibility of exotic proton decay modes such as $p\to\pi^{+}\bar{N}^{0}$, whose viability depends on the pNGB spectrum discussed in \cref{sec:pNGB}.} The Super-Kamiokande limit $\tau(p\to e^+\pi^0) > 1.6\times 10^{34}~{\rm yr}$ then yields\footnote{In line with the NDA estimates of prebaryon bound state contributions to the quark and charged lepton Yukawas in \cref{sec:masses} we do not include loop factors when replacing two prebaryon legs with the corresponding composite scalar. 
  In this counting, the baryon number violating dimension 6 operator $(u_L d_L)(u_Re_R)/M_{udql}^2$ is suppressed by $M_{udql}=N^{3/2} N_c \Lambda_\text{pre}^2/c_\phi (C_8 m_{7/6} m_{33})^{1/2}$, which differs by a factor of $16\pi^2$ from the NDA estimate in \cite{Assi:2022jwg}.
 Whether or not the inclusion of this loop factor gives a better estimate of the unknown nonperturbative matrix elements is an open question, however, it should be applied consistently in both the formation of SM Yukawa interactions and in proton decay bounds. }
\begin{equation}
\frac{\Lambda_{\rm pre}}{|C_8|^{1/4}} \;>\; 1.3\times 10^{5}~{\rm TeV}\;\left(\frac{m_{7/6}\, m_{33}}{{\rm TeV}^{2}}\right)^{1/4},
\label{eq:proton_bound}
\end{equation}
with $C_8$ an order-unity matching coefficient and $m_{7/6}=y_{7/6}\langle\phi_{7/6}\rangle$, $m_{33}=y_{33}\langle\phi_{33}\rangle$ the masses of the composite vectorlike quarks $\mathcal{Q}_{7/6}$ and $\mathcal{Q}_{33}$, respectively. The bound therefore depends on the composite-scalar vevs that fix the vector-like quark spectrum, as well as on the value of the unknown Wilson coefficient $C_8$, which encodes the $\SUp$ dynamics. 

Varying vectorlike quark masses from $2$ to $30\,$TeV, and $C_8$ in the range $10^{-5}$ to $10$~\cite{Assi:2022jwg}, the proton lifetime limit implies a lower bound on the compositeness scale $\Lambda_{\rm pre}$ in the $10^{4}$--$10^{6}$~TeV range. This is suggestive that precision flavour measurements may provide complementary constraints on $\Lambda_{\rm pre}$, which we discuss in the rest of this section. 

\subsection{Dipole operators}

Attaching a photon to any of the preon lines that carry a nonzero electromagnetic charge, $\psi_{Q,U,D, L,E}$, in the diagrams for the quark and charged lepton Yukawas, Figs. \ref{fig:quark_diagrams}, \ref{fig:lepton_diagram}, and \ref{fig:lepton_nonlocal}, gives rise to magnetic and electric dipole moment operators
\beq
\label{eq:Leff:dipole}
{\cal L}_\text{eff}\supset \frac{e \big(C_{f}^{\gamma\prime}\big)_{ij} \langle H_{u,d}\rangle}{\Lambda_\text{pre}^2} (\bar f_{i} \sigma_{\mu\nu} P_R f_j)F^{\mu\nu} + \text{h.c.},
\eeq
where $f=u,d,\ell$ are the SM fermions, and $i,j=1,2,3$ are generation indices, with $\langle H_u\rangle=v_u/\sqrt 2$ the Higgs vev appearing for $f=u$, and $\langle H_d\rangle=v_d/\sqrt{2}$ for $f=d, \ell$. 

The Wilson coefficients have a very similar structure in terms of the spurion $\lambda, \lambda'$ insertions, as the SM Yukawa matrices, \cref{eq:quark_yukawa,eq:lepton_yukawa},
\begin{equation}
    \begin{split}
    \label{eq:dipole:ops}
   \big( C_{f}^{\gamma\prime}\big)_{ij} =& F_{f}^{\prime \gamma} 
        \lambda'_{43}\lambda_{ij}^{\prime *} +
      F_{f}^\gamma 
        \lambda_{43}\lambda_{ij}^{*}
    + \frac{1}{16\pi^2} \Big[ G_{f}^\gamma \lambda_{3a} \lambda_{ia}^* \lambda_{4b}\lambda_{jb}^*+ I_{f}^\gamma \lambda'_{3a} \lambda_{ia}^* \lambda'_{4b}\lambda_{jb}^*
    \\
    &\qquad\qquad\qquad\qquad\qquad\qquad\quad + J_{f}^\gamma \lambda_{3a} \lambda_{ia}^{\prime *} \lambda_{4b}\lambda_{jb}^{\prime*}+ K_{f}^\gamma \lambda'_{3a} \lambda_{ia}^{\prime*} \lambda'_{4b}\lambda_{jb}^{\prime *}\Big],
    \end{split}
\end{equation}
where we expect the non-perturbative functions $F_f^{(\prime)\gamma}, G_f^\gamma,\ldots$ to be ${\mathcal O}(1)$ (similar to the magnetic moments of the proton and neutron). That is, we expect these nonperturbative functions to be of similar size as the ones for quark and lepton Yukawas, but also not to be exactly the same, $F_f^{(\prime)\gamma}\neq F_f^{(\prime)}, G_f^\gamma \neq G_f,\ldots$. In the numerics we set all the nonperturbative parameters in \cref{eq:dipole:ops} to 1 for definiteness, i.e., $F_f^{\prime \gamma}=1$, \ldots. 

In the mass basis, the dipole operator Wilson coefficients are given by,
\beq
C_f^\gamma = U_f^\dagger\, \hat C_f^{\gamma\prime}\, V_f,
\eeq
and thus in general have both diagonal and off-diagonal entries, i.e., we have $\big(C_f^\gamma\big)_{ij}\ne0$ also for $i\ne j$. That is, while parametrically the dipole and the corresponding Yukawas are similar in size, $C_f^\gamma\sim Y_f$, they are not diagonalized in the same basis, and thus we expect that in the mass basis the dipole Wilson coefficient $C_f^\gamma$ still have both diagonal and off-diagonal components of the following parametric size
\beq
\label{eq:Cf:parametric}
C_f^\gamma
\sim \begin{pmatrix}
  \kappa^{4} & \kappa^{3} & \kappa^{3}
  \\
 \kappa^{3} & \kappa^{2} & \kappa^{2}
 \\
 \kappa^{3} & \kappa^{2} & 1
 \end{pmatrix}~, 
 \qquad f=u,d,\ell~.
\eeq

Focusing first on the leptonic dipoles, 
in our numerical benchmark several of the entries are numerically enhanced, 
\beq
\label{eq:dipole:lepton}
C_\ell^\gamma\big|_\text{bench.}
\sim \begin{pmatrix}
  \kappa^{3} & \kappa^{3} & \kappa^{2}
  \\
 \kappa^{3} & \kappa^{2} & \kappa^{2}
 \\
 \kappa^{2} & \kappa^{} & 1
 \end{pmatrix}.
\eeq
Since the dipole operator \cref{eq:Leff:dipole} is chirality flipping in many flavour models its size is proportional to geometric mean of appropriate lepton Yukawas, ${\sqrt{m_{\ell_i} m_{\ell_j}}}/{v_d}$, which is parametrically given by
\beq
\label{eq:geometric:mean:lepton}
\frac{\sqrt{m_{\ell_i} m_{\ell_j}}}{v_d}
\sim 
\begin{pmatrix}
  \kappa^{5} & \kappa^{7/2} & \kappa^{5/2}
  \\
 \kappa^{7/2} & \kappa^{2} & \kappa
 \\
 \kappa^{5/2} & \kappa & 1
 \end{pmatrix}.
\eeq
We see that in the two-scalar preon model (with the assumed benchmark textures), the charged lepton transition dipole moments are roughly of the geometric mean size in \cref{eq:geometric:mean:lepton}, though also with differences at the level factors of a few. 
Interestingly, the electric dipole moment is expected to be enhanced relative to the naive $m_e/v_d$ chiral suppression. 

The benchmark values of the quark dipole moments are numerically of the following size
\beq
C_u^\gamma\big|_\text{bench.}
\sim \begin{pmatrix}
  \kappa^{4} & \kappa^{2} & \kappa^{2}
  \\
 \kappa^{4} & \kappa^{2} & \kappa^{2}
 \\
 \kappa^{3} & \kappa^{2} & 1
 \end{pmatrix},
 \qquad
 C_d^\gamma\big|_\text{bench.}
\sim \begin{pmatrix}
  \kappa^{4} & \kappa^{3} & \kappa^{2}
  \\
 \kappa^{3} & \kappa^{2} & \kappa^{2}
 \\
 \kappa^{2} & \kappa^{2} & 1
 \end{pmatrix}.
\eeq
It is interesting to note that $b\to s\gamma$ and $b\to d\gamma$ are both ${\mathcal O}(\kappa^2)$ in the benchmark.

\paragraph{Constraints from transition dipole operators.}
The off-diagonal dipole operators lead to new physics contribution to $b\to s\gamma$, correcting the SM decay rates that get induced at 1-loop. They also generate FCNC decays of charged leptons, such as $\mu \to e\gamma$, $\tau\to e\gamma, \mu\gamma$ with branching ratios \cite{Calibbi:2018rzv}
\beq
\frac{\text{Br}(\ell_i\to \ell_j \gamma)}{\text{Br}(\ell_i\to \ell_j \nu_i \bar \nu_j)}=3 \biggr(\frac{4\pi e \langle H_d\rangle}{G_F m_{\ell_i} \Lambda_\text{pre}^2}\biggr)^2 \big(\overline{C_\ell^\gamma}\big)_{ji}^2,
\eeq
where we ignored the masses of the final state particles for simplicity and defined 
\beq
\big(\overline{C_\ell^\gamma}\big)_{ji}=\Big\{ \big|\big(C_\ell^\gamma\big)_{ji}\big|^2+\big|\big(C_\ell^\gamma\big)_{ij}\big|^2\Big\}^{1/2}.
\eeq
Present experimental constraints on these decays translate into the following bounds 
\begin{align}
&\mu\to e\gamma\text{~\cite{MEGII:2025gzr}}: & \Lambda_\text{pre}> 9.0 \cdot 10^{2}~{\rm TeV} \times \biggr(\frac{1.5 \cdot 10^{-13}}{\text{Br}(\mu\to e\gamma)}\biggr)^{1/4} \biggr(\frac{\big(\overline{C_\ell^\gamma}\big)_{12}}{\bench{7.7 \cdot 10^{-3}}} \cdot\frac{\cos\beta}{\bench{0.044}}\biggr)^{1/2},
\\
&\tau\to e\gamma \text{~\cite{BaBar:2009hkt}}: & \Lambda_\text{pre}> 23~{\rm TeV} \times \biggr(\frac{3.3 \cdot 10^{-8}}{\text{Br}(\tau\to e\gamma)}\biggr)^{1/4} \biggr(\frac{\big(\overline{C_\ell^\gamma}\big)_{13}}{\bench{9.1 \cdot 10^{-2}}} \cdot\frac{\cos\beta}{\bench{0.044}}\biggr)^{1/2},
\\
&\tau\to \mu\gamma \text{~\cite{Belle:2021ysv}}: & \Lambda_\text{pre}> 35~{\rm TeV} \times \biggr(\frac{4.2 \cdot 10^{-8}}{\text{Br}(\tau\to \mu\gamma)}\biggr)^{1/4} \biggr(\frac{\big(\overline{C_\ell^\gamma}\big)_{23}}{\bench{0.25}} \cdot\frac{\cos\beta}{\bench{0.044}}\biggr)^{1/2},
\end{align}
where for the numerical values we used the current $90\%$CL experimental upper bounds on the branching ratios, and the benchmark values of the parameters. The future sensitivity, corresponding to $\text{Br}(\mu\to e\gamma)< 6 \cdot 10^{-14}$ projected sensitivity at MEG-II \cite{MEGII:2018kmf} after three years of running is $\Lambda_\text{pre}>1.1\cdot 10^3$\,TeV, and $\Lambda_\text{pre}>31 (55)\,$TeV at Belle-II with $50\,\text{ab}^{-1}$ integrated luminosity from $\text{Br}(\tau\to e(\mu)\gamma)< 9.0 (6.9) \cdot 10^{-9}$~\cite{Banerjee:2022vdd}.

The bounds on deviations of experimental measurements for $b\to s\gamma$ transitions from the SM predictions for $\text{Br}(B\to X_s\gamma), \text{Br}(B^{+,0}\to K^{*+,0}\gamma), \text{Br}(B_s\to \phi \gamma)$, $A_{\Delta \Gamma} (B_s\to \phi \gamma)$, $S_{K*\gamma}$ give \cite{Straub:2018kue,Straub:2026flavio} 
\beq
\frac{\big|\big(C_{d}^\gamma\big)_{23 (32)}\big| \langle H_{d}\rangle}{\Lambda_\text{pre}^2}< 6 (8)\times 10^{-9} \,\text{GeV}^{-1},
\eeq
which translates to 
\beq
b\to s\gamma:\qquad \Lambda_\text{pre}>14\,\text{TeV}\times \biggr(\frac{\big|\big(C_{d}^\gamma\big)_{32}\big|}{\bench{0.16}}\biggr)^{1/2}.
\eeq
For the numerical value we used the stronger of the two bounds, $6\times 10^{-9}\,\text{GeV}^{-1}$; using instead $8\times 10^{-9}\,\text{GeV}^{-1}$ would give $\Lambda_\text{pre}>\bench{9.6}$\,TeV.
The decay $b\rightarrow d \gamma$ probes similar scales, but both decays are subdominant (by orders of magnitude) to other observables. 

\paragraph{Electric and magnetic dipole moments.}
The flavour diagonal dipole operators in \cref{eq:Leff:dipole} give rise to magnetic, $\mu_f$, and electric dipole moments, $d_f$, for fermion $f$, \cite{Pospelov:2025vzj,Chupp:2017rkp}
\beq 
{\cal L}_\text{diag}=-\frac{1}{2} \mu_f (\bar f \sigma^{\mu\nu} f) F_{\mu\nu}-\frac{i}{2} d_f (\bar f \sigma^{\mu\nu} \gamma_5 f) F_{\mu\nu}.
\eeq
In terms of Wilson coefficients in \cref{eq:Leff:dipole} we have 
\beq
\mu_{f_i}=-\frac{2 e \langle H_{u,d}\rangle}{\Lambda_\text{pre}^2}\Re\big( C_{f}^\gamma\big)_{ii}, \qquad 
d_{f_i}=-\frac{2 e \langle H_{u,d}\rangle}{\Lambda_\text{pre}^2}\Im\big( C_{f}^\gamma\big)_{ii}.
\eeq
In theories with unsuppressed CP violating phase, as is the case for the two scalar preon model, the bounds on electric dipole moments lead to much more stringent constraints. We thus focus exclusively on these. 

The bounds on the electron EDM from ${\rm Hf\hspace{1pt}F}^+$, $d_e < 4.1 \cdot 10^{-30}\,e\,{\rm cm}$~\cite{Roussy:2022cmp}, and on the muon EDM, $d_\mu <1.4 \cdot 10^{-19}e\,$cm~\cite{Muong-2:2008ebm}, give
\begin{align}
e\text{-EDM}:&\qquad \Lambda_\text{pre}>9.5 \cdot 10^{3}\, \text{TeV} \times \biggr(\frac{\big|\Im \big({C_\ell^\gamma}\big)_{11}\big|}{\bench{1.2 \cdot 10^{-3}}} \cdot\frac{\cos\beta}{\bench{0.044}}\biggr)^{1/2},
\\
\mu\text{-EDM}:&\qquad \Lambda_\text{pre}>0.15 \,\text{TeV} \times \biggr(\frac{\big|\Im \big({C_\ell^\gamma}\big)_{22}\big|}{\bench{1.3 \cdot 10^{-2}}} \cdot\frac{\cos\beta}{\bench{0.044}}\biggr)^{1/2}.
\end{align}If the bound on electron EDM gets lowered to $d_e < 10^{-31}\,e\,{\rm cm}$, as targeted by next generation of experiments \cite{Alarcon:2022ero}, this would imply a reach $\Lambda_\text{pre}>6.1 \cdot 10^{4}\,$TeV. In the next two decades improvements of six orders may even be possible \cite{Alarcon:2022ero}; $d_e \lesssim 10^{-34}\,e\,{\rm cm}$ would imply sensitivity to $\Lambda_\text{pre}\gtrsim 2 \cdot 10^{6}\,$TeV, significantly surpassing the current typical proton bounds, \eqref{eq:proton_bound}. For $\mu$-EDM the muEDM experiment at PSI targets sensitivity, of $d_\mu <6 \cdot 10^{-23}e\,$cm \cite{Sakurai:2022tbk}, which would imply $\Lambda_\text{pre}>8\,\text{TeV}$.

The neutron EDM receives contributions from both up and down quark EDMs \cite{Pospelov:2025vzj}, 
\beq
\label{eq:dn}
d_n=g_T^d d_d+g_T^u d_u+\cdots,
\eeq
where $g_T^u = 0.784(30)$, $g_T^d=-0.204(15)$~\cite{FlavourLatticeAveragingGroupFLAG:2021npn,Gupta:2018lvp,Haxton:2024lyc}. The ellipses in 
\cref{eq:dn} denote additional contributions from chromo-EDMs and the CP-odd three gluon Weinberg operator, which are parametrically of similar size as the quark EDM ones. 
In the limit of either $d_d$ or $d_u$ dominance, the bound on neutron EDM, $d_n<1.8 \cdot 10^{-26}e\,$cm implies
\beq
\frac{\big|\big(C_{u(d)}^\gamma\big)_{11}\big| \langle H_{u(d)}\rangle}{\Lambda_\text{pre}^2}< 0.6 (2.2)\times 10^{-12} \,\text{GeV}^{-1}.
\eeq
For our numerical benchmark both up-quark and down-quark EDMs give comparable contributions. Keeping both, gives the following bound on the preon scale
\beq
d_n:\qquad \Lambda_\text{pre}>240 \,\text{TeV}.
\eeq
which is much weaker than that obtained from the electron EDM. Next generation neutron EDM experiments will target $d_n\lesssim 10^{-27}e\,$cm sensitivity \cite{n2EDM:2021yah,TUCAN:2025rjm,Wurm:2019yfj}, which would translate to $\Lambda_\text{pre}\gtrsim10^3 \,\text{TeV}$.

\paragraph{Muon conversion on nuclei.}
In two-scalar preon model the leading contribution to $\mu \to e$ conversion is due to the transition dipole moment, \cref{eq:Leff:dipole}, where the photon attaches to the protons inside the nucleus. In general, $\mu\to e$ conversion also receives contributions from four-fermion operators of the form $(\bar e \gamma^\nu P_{L,R} \mu)(\bar q \gamma_\nu q)$, with $q=u,d$, see \cref{sec:semileptonic}. These contributions are, however, highly suppressed in our benchmark. 
That is, while $[C_\ell^\gamma]_{12}\sim [C_\ell^\gamma]_{21} \sim \kappa^3$, cf. \cref{eq:Cf:parametric,eq:dipole:lepton}, the semi-leptonic four fermion operators due to tree level ${\mathcal A}, {\mathcal A'}$ exchanges are instead ${\mathcal O}(\kappa^7)$ suppressed, with loop level contributions of order ${\mathcal O}(\kappa^3/16\pi^2)$. These contributions are numerically small, and can therefore be safely neglected. 

The $\mu\to e$ conversion rate is conventionally normalized to the SM muon capture rate $\Gamma_{\rm capt}=\Gamma(\mu^-+(A,Z)\to \nu_\mu+(A,Z-1))$, 
\begin{equation}
    {\rm CR}(\mu \to e) = \frac{\Gamma(\mu\to e)}{\Gamma_{\rm capt}}.
\end{equation}
For aluminum, $\Gamma_{\rm capt}|_\text{Al}=6.982(12)\cdot 10^5\,\text{s}^{-1}$, while for titanium, $\Gamma_{\rm capt}|_\text{Ti}=2.592(11)\cdot 10^6\,\text{s}^{-1}$~\cite{Suzuki:1987jf}. The $\mu\to e$ conversion rate is given by \cite{Haxton:2024lyc,Haxton:2022piv}
\beq
\Gamma(\mu \to e)=\frac{q_\text{eff}^2}{2\pi} \big|\phi_{1s}^{Z_\text{eff}}(0)\big|^2 |c_1^p|^2 W_{MM}^{pp} (q_\text{eff}^2),
\eeq
where we only included the coherently enhanced contributions due to transition dipole moments,
\beq
    c_1^p=\frac{e^2 \langle H_d\rangle}{q_\text{eff} \Lambda_\text{pre}^2 }\Big( [C_\ell^\gamma]_{12} + [C_\ell^\gamma]_{21}^* \Big)~.
\eeq
Note that the effective momentum exchange $q_\text{eff}$ in the prefactor, cancels against $c_1^p$, with the remaining dependence only in the nuclear response functions. Numerically, $q_\text{eff}|_\text{Al}=110.81\,$MeV for aluminum, and $q_\text{eff}|_\text{Ti}=112.43\,$MeV for titanium \cite{Haxton:2022piv}. The nuclear response function for coherent conversion due to couplings to protons, $W_{MM}^{pp}=\frac{1}{4}\big(W_{MM}^{00}+W_{MM}^{11}+2 W_{MM}^{01}\big)\sim {\mathcal O}(Z^2)$, is $W_{MM}^{pp} (q_\text{eff}^2)|_\text{Al}\simeq 131$ for aluminum and $W_{MM}^{pp} (q_\text{eff}^2)|_\text{Ti}\simeq 585$ for titanium \cite{Haxton:2022piv}. The value of the wave function at the origin for the muon in $1s$ state is
\beq
\phi_{1s}^{Z_\text{eff}}(0)=\frac{1}{\sqrt \pi} (Z_\text{eff} \alpha m_\mu)^{3/2},
\eeq
with the effective charges $Z_\text{eff}|_\text{Al}=11.31$, $Z_\text{eff}|_\text{Ti}=16.66$~\cite{Haxton:2022piv}. 

For the numerical benchmark, the bound presented by SINDRUM II at a conference $ \text{CR}(\mu\to e)|_\text{Ti}<6.1\cdot 10^{-13}$ \cite{Wintz1998} (to be compared with the published bound $ \text{CR}(\mu\to e)|_\text{Ti}<4.3\cdot 10^{-12}$ \cite{SINDRUMII:1993gxf}) implies the following constraint on $\Lambda_\text{pre}$ for our benchmark
\begin{align}
&\mu\to e\big|_\text{Ti} : & \Lambda_\text{pre}> 220\,{\rm TeV} \times \biggr(\frac{6.1\cdot 10^{-13}}{\text{CR}(\mu\to e)|_\text{Ti}}\biggr)^{1/4} \biggr(\frac{\big|[C_\ell^\gamma]_{12} + [C_\ell^\gamma]_{21}^*\big|}{9.3 \cdot 10^{-3}} \cdot\frac{\cos\beta}{0.044}\biggr)^{1/2},
\\
&\mu\to e\big|_\text{Al}: & \Lambda_\text{pre}> 2.5 \cdot 10^3\,{\rm TeV} \times \biggr(\frac{10^{-17}}{\text{CR}(\mu\to e)|_\text{Al}}\biggr)^{1/4} \biggr(\frac{\big|[C_\ell^\gamma]_{12} + [C_\ell^\gamma]_{21}^*\big|}{9.3 \cdot 10^{-3}} \cdot\frac{\cos\beta}{0.044}\biggr)^{1/2}.
\end{align}
The second line shows the typical expected future reach for muon conversion on aluminum target at Mu2e \cite{Mu2e-II:2022blh} and COMET \cite{Moritsu:2022lem,Fujii:2023vgo}.

\paragraph{Three-body lepton decays:}
Focusing on the FCNC decays of the form $\ell_1\to 3 \ell_2$, the branching ratios for these are given by, in the dipole operator dominance approximation \cite{Arganda:2005ji}, 
\beq
\frac{\text{Br}(\ell_1\to 3 \ell_2)}{\text{Br}(\ell_1\to \ell_2 \gamma)}=\frac{\alpha}{3\pi}\biggr(\log \frac{m_{\ell_1}^2}{m_{\ell_2}^2}-\frac{11}{4}\biggr).
\eeq
The current bounds ${\rm BR}(\mu\to 3e) < 1.0\times 10^{-12}$~\cite{SINDRUM:1987nra}, ${\rm BR}(\tau\to 3e) < 2.7\times 10^{-8}$ \cite{Belle-II:2025urb}, ${\rm BR}(\tau\to 3\mu) < 1.9\times 10^{-8}$ \cite{Belle-II:2024sce} yield
\begin{align}
&\mu\to 3e \text{~\cite{SINDRUM:1987nra}}: & \Lambda_\text{pre}> \bench{1.6\cdot10^{2}}~{\rm TeV} \times \biggr(\frac{1.0 \cdot 10^{-12}}{\text{Br}(\mu\to 3 e)}\biggr)^{1/4} \biggr(\frac{\big(\overline{C_\ell^\gamma}\big)_{12}}{\bench{7.7 \cdot 10^{-3}}} \cdot\frac{\cos\beta}{\bench{0.044}}\biggr)^{1/2},
\\
&\tau\to 3 e \text{~\cite{Belle-II:2025urb}}: & \Lambda_\text{pre}> \bench{7.6}~{\rm TeV} \times \biggr(\frac{2.7 \cdot 10^{-8}}{\text{Br}(\tau\to 3 e)}\biggr)^{1/4} \biggr(\frac{\big(\overline{C_\ell^\gamma}\big)_{13}}{\bench{9.1 \cdot 10^{-2}}} \cdot\frac{\cos\beta}{\bench{0.044}}\biggr)^{1/2},
\\
&\tau\to 3 \mu \text{~\cite{Belle-II:2024sce}}: & \Lambda_\text{pre}> \bench{9.3}~{\rm TeV} \times \biggr(\frac{1.9 \cdot 10^{-8}}{\text{Br}(\tau\to 3 \mu )}\biggr)^{1/4} \biggr(\frac{\big(\overline{C_\ell^\gamma}\big)_{23}}{\bench{0.25}} \cdot\frac{\cos\beta}{\bench{0.044}}\biggr)^{1/2},
\end{align}
The upcoming Mu3e experiment projects sensitivity to ${\rm BR}(\mu\to 3e)\sim 10^{-16}$~\cite{Mu3e:2020gyw}, which would then lead to a sensitivity to $\Lambda_\text{pre}> \bench{1.6 \cdot 10^{3}}~{\rm TeV}$, while Belle-II is expected to reach sensitivity to ${\rm BR}(\tau\to 3e(3\mu))\sim 4.7 (3.6)\cdot10^{-10}$~\cite{Banerjee:2022vdd}, which would imply sensitivity to $\Lambda_\text{pre}> \bench{21}\,(\bench{25})~{\rm TeV}$.

\subsection{Semi-leptonic four-fermion operators}
\label{sec:semileptonic}
The diagrams involving exchanges of $\mathcal{A}$ and $\mathcal{A}'$ scalars lead to FCNC operators with four SM fermions on the external legs. We start with the semi-leptonic operators. Due to the assumed flavor structure of the $\lambda, \lambda'$ couplings, \cref{eq:lambda_texture}, we need to include both tree and one-loop exchanges of ${\mathcal A}, {\mathcal A}'$, cf. \cref{fig:semileptonic_topology}. In the mass basis, the resulting effective Lagrangian is given by 
\beq
\begin{split}
\label{eq:semileptonic:both}
        \mathcal{L}_{\rm semilep}= \frac{1}{\Lambda_{\rm pre}^2}&\Big\{ [C_{LL}^{\ell q}]_{ij;kl} (\bar{\ell}_i \gamma^\mu P_L \ell_j)(\bar{q}_k \gamma_\mu P_L q_l) - [C_{LR}^{\ell q}]_{ij;lk} (\bar{\ell}_i \gamma^\mu P_L \ell_j)(\bar{q}_k \gamma_\mu P_R q_l)
        \\
         - & [C_{RL}^{\ell q}]_{ji;kl} (\bar{\ell}_i \gamma^\mu P_R \ell_j)(\bar{q}_k \gamma_\mu P_L q_l) + [C_{RR}^{\ell q}]_{ji;lk} (\bar{\ell}_i \gamma^\mu P_R \ell_j)(\bar{q}_k \gamma_\mu P_R q_l) \Big\}+ {\rm h.c.},
\end{split}        
\eeq
where the summation over the generation indices $i,j, k, l$ is implicit, 
and $q$ is either $u$ or $d$. The coefficients of different four-fermion operators are a sum of tree-level and one-loop contributions
\beq
\label{eq:sum:CAB}
[C_{AB}^{\ell q}]_{ij;kl}=[C_{AB}^{\ell q}]_{ij;kl}^\text{tree}+[C_{AB}^{\ell q}]_{ij;kl}^\text{loop},
\eeq
with the contributions due to tree level ${\mathcal A}, {\mathcal A}'$ exchanges given by,
\beq
\label{eq:CAB:lq:tree}
[C_{AB}^{\ell q}]_{ij;kl}^\text{tree}=F_4 [\lambda_{\ell_A q_B}]_{jl}[\lambda_{\ell_A q_B}^*]_{ik} +F_4' [\lambda_{\ell_A q_B}']_{jl}[\lambda_{\ell_A q_B}^{\prime*}]_{ik}, 
\eeq
and due to one loop box exchanges by 
\beq
\begin{split}
\label{eq:CAB:loop}
 [C_{AB}^{\ell q}]_{ij;kl}^\text{loop}=\frac{1}{16\pi^2} \Big\{& G_4
        [\lambda_{\ell_A}^{(2)}]_{ji} [\lambda_{q_B}^{(2)}]_{lk}       
        + G_4'
        [\lambda_{\ell_A}^{\prime(2)}]_{ji} [\lambda_{q_B}^{(2)}]_{lk}
       \\
      +& G_4''
        [\lambda_{\ell_A}^{(2)}]_{ji} [\lambda_{q_B}^{\prime (2)}]_{lk}
+ G_4'''
        [\lambda_{\ell_A}^{\prime(2)}]_{ji} [\lambda_{q_B}^{\prime (2)}]_{lk}\Big\}.
 \end{split}
\eeq
 The nonperturbative functions, $F_4^{(\prime)}, G_4^{(\prime)(\prime)(\prime)}$, are expected to be ${\mathcal O}(1)$, and are set to $F_4^{(\prime)}=G_4^{(\prime)(\prime)(\prime)}=1$ in the numerics below. 
Note that for simplicity we ignore small electroweak corrections that distinguish at one loop between $q=u$ and $q=d$ values for these nonperturbative factors. 

The FCNC structure resides in the rotated combinations of ${\mathcal A}, {\mathcal A}'$ couplings, e.g., for left-handed currents
\begin{align}
\label{eq:lambdaff'}
\lambda_{f_L f'_L}&=U_{f}^T[\lambda]_{3\times 3} U_{f'}, &\lambda'_{f_L f'_L}&=U_{f}^T[\lambda']_{3\times 3} U_{f'},
\\
\lambda_{f_L }^{(2)}&=U_{f}^T[\lambda \lambda^\dagger]_{3\times 3} U_{f}^*, &\lambda_{f_L}^{\prime (2)}&=U_{f}^T[\lambda' \lambda^{\prime \dagger}]_{3\times 3} U_{f}^*,
\end{align}
with $U_f$ the left-handed rotation matrices in SVD, cf. \cref{eq:Yud:rotate,eq:Yell:rotate}. The $[\lambda]_{3\times 3}$, $[\lambda']_{3\times 3}$, $[\lambda \lambda^\dagger]_{3\times 3}$, and $[\lambda'\lambda^{\prime \dagger}]_{3\times 3}$ are the $3\times 3$ blocks of the corresponding (products of) spurion matrices, i.e., with the fourth row and column removed. The expressions for couplings involving right-handed fermions $f_R, f_R'$ are obtained by replacing $U_f\to V_f^*$, etc. Explicitly, $\lambda_{f_R f'_L}=V_{f}^\dagger [\lambda]_{3\times 3} U_{f'}$, $\lambda_{f_L f'_R}=U_{f}^T [\lambda]_{3\times 3} V_{f'}^*$, and $\lambda_{f_R f'_R}=V_{f}^\dagger [\lambda]_{3\times 3} V_{f'}^*$, and similarly for $\lambda'$, while the spurions entering at 1-loop level are $\lambda_{f_R}^{(2)}=V_{f}^\dagger [\lambda \lambda^\dagger]_{3\times 3} V_{f}$, and $\lambda_{f_R}^{\prime (2)}=V_{f}^\dagger [\lambda' \lambda^{\prime \dagger}]_{3\times 3} V_{f}$. The negative signs for the LR and RL terms in \cref{eq:semileptonic:both} arise from identities when converting between two- and four-component spinors \cite{Dreiner:2008tw}. 

The semi-leptonic operators in \cref{eq:semileptonic:both} give rise to FCNC that involve both quark and lepton currents. Below we discuss bounds on preon compositeness scale $\Lambda_\text{pre}$ placed by searches for $K_L\to \mu e$ and $K\to \pi \nu\nu$ decays. Since dipole couplings dominate over direct four-fermion couplings in muon conversion after fitting to the observed lepton masses, we do not discuss muon conversion on nuclei here.

\begin{figure}[t!]
  \begin{center}
 \includegraphics[width=0.95\textwidth]{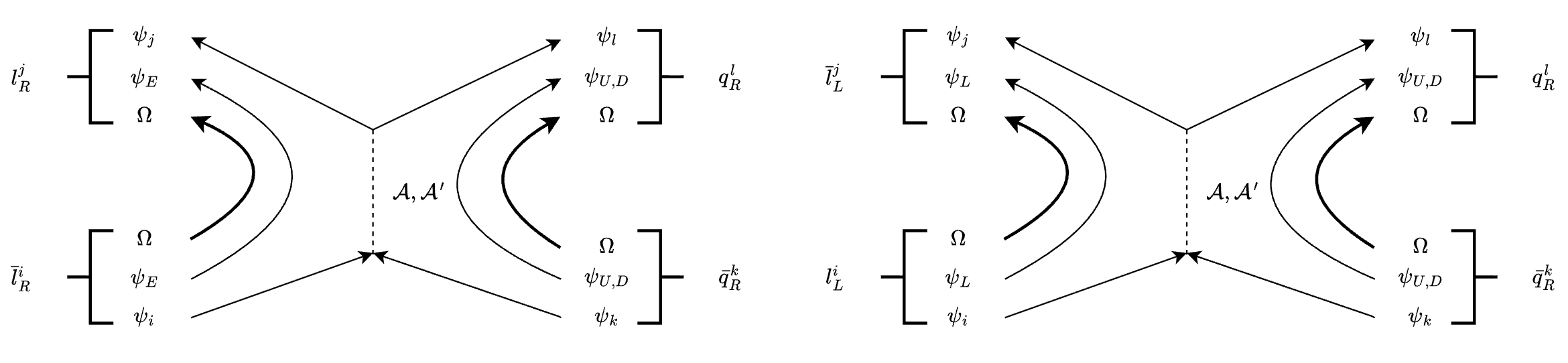}
 \\
 \includegraphics[width=0.95\textwidth]{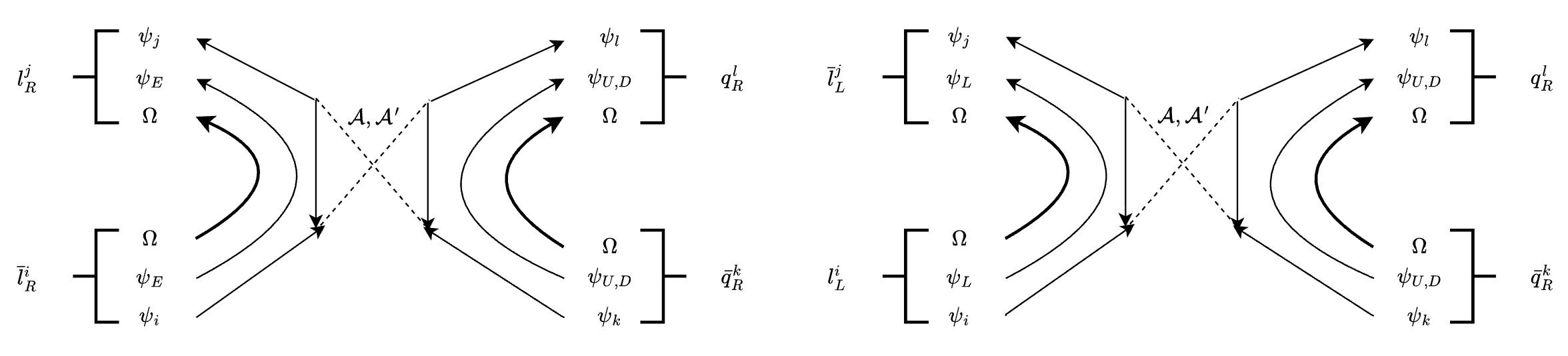}
   \caption{Contributions to semileptonic four-fermion operator from tree level (top row) or loop level (bottom row) $\mathcal{A}$ or $\mathcal{A}'$ exchanges, coupling lepton currents $\bar{\ell}'\gamma_\mu P_{L,R} \ell$ to right-handed down quark current $\bar{q}' \gamma^\mu P_R q$. Similar contributions giving rise to operators with right-handed up quark currents (left-handed quark currents) follow from $\psi_D\to \psi_{U(Q)}$ replacements. }
  \label{fig:semileptonic_topology}
  \end{center}
\end{figure}

\paragraph{Lepton flavour violating $K_L\to e^\pm\mu^\mp$ decays.} 
These decays are suppressed by the neutrino masses in the SM, and thus completely negligible. In the two scalar preon model they are induced by the semileptonic four-fermion operators with hadronic axial currents,
\begin{equation}
\mathcal L_{\rm eff}^{\Delta S=1}
\supset
\frac{1}{\Lambda_{\rm pre}^2}\,
\big(C_{L;sd}^{12}(\bar{e}_L \gamma_\alpha \mu_L) + C_{R;sd}^{1 2} (\bar{e}_R\gamma_\alpha \mu_R) + e\leftrightarrow \mu, 1\leftrightarrow 2\big)\,
(\bar s \gamma^\alpha \gamma_5 d)+ {\rm h.c.},
\label{eq:KLemu-Leff-general}
\end{equation}
where the Wilson coefficients are given by  
\begin{align}
\label{eq:CL:ij}
C_{L;sd}^{ij}&=-\frac{1}{2}\Big( [C_{LL}^{\ell d}]_{ij;21}+[C_{LR}^{\ell d}]_{ij;12}\Big),
  \\
  \label{eq:CR:ij}
C_{R;sd}^{ij}&=+\frac{1}{2} \Big( [C_{RL}^{\ell d}]_{ji;21}+[C_{RR}^{\ell d}]_{ji;12}\Big).  
 \end{align}
The tree level contributions have parametric size $C_{L(R);sd}^{12}, C_{L(R);sd}^{21}\sim {\mathcal O}(\kappa^6)$, while loop contributions are $ {\mathcal O}(\kappa^2/16\pi^2)$, and are thus numerically larger. The reason for this enhancement is that in the loops the summation over virtual states includes summations over all four flavors, $a=1,\ldots,4$, which occurs at 1-loop level for the first time. 
 
 The decay widths for $K_L\to e^\pm\mu^\mp$ decays are
 \beq
 \Gamma(K_L\to e^-\mu^+)= \Gamma(K_L\to e^+\mu^-)=\frac{1}{64\pi}\biggr(\frac{f_K m_\mu}{\Lambda_\text{pre}^2}\biggr)^2\Big(\big|\bar C_{L;sd}^{e\mu}\big|^2+\big|\bar C_{R;sd}^{e\mu}\big|^2\Big)\Big(1-\frac{m_\mu^2}{m_K^2}\Big)^2,
 \eeq
 with $f_K=155\,$MeV~\cite{FlavourLatticeAveragingGroupFLAG:2024oxs} the kaon decay constant and 
 \beq
 \bar C_{L(R);sd}^{e\mu}=C_{L(R);sd}^{12*}- C_{L(R);sd}^{21}.
 \eeq
 Using ${\rm BR}(K_L\to e^\pm\mu^\mp)<4.7\times10^{-12}$ \cite{BNL:1998apv} we obtain
\begin{equation}
\Lambda_{\rm pre}
\ \gtrsim\
\bench{2.2}\,{\rm TeV}\times
\biggr(\frac{ \sqrt{\big|\bar C_{L;sd}^{e\mu}\big|^2+\big|\bar C_{R;sd}^{e\mu}\big|^2}}{\bench{2.4 \cdot 10^{-5}}} \biggr)^{1/2},
\label{eq:KL-Lpre-final}
\end{equation}where we used the benchmark numerical values for $ \bar C_{L(R);sd}^{e\mu}$ in the above estimate. 

\paragraph{The $K\to \pi \nu\bar\nu$ decays.}
The $K_L \to \pi^0 \nu\bar{\nu}$ and $K^+ \to \pi^+ \nu\bar{\nu}$ decays arise in the SM at one loop order, via 
$Z$-penguin and box diagrams. The SM predictions are exceptionally clean, making these ``golden modes'' for testing new physics \cite{Buras:2015qea,Buras:2024ewl,Brod:2021hsj}.

In the two scalar preon model the $s\to d\nu\bar \nu$ transitions are generated by the same semi-lepton operators as for charged leptons, \cref{fig:semileptonic_topology}, but now with neutrinos on the external legs. Since in our setup there are no light right-handed neutrinos, the neutrino currents are necessarily left-handed. Furthermore, only the vector part of the quark current contributes to $K\to \pi \nu\bar \nu$ decays, so that the relevant part of the $\Delta S=1$ Lagrangian is given by
\beq
\mathcal L_{\rm eff}^{\Delta S=1} \supset
\sum_{ij} \hat C_{ij}^{\nu\nu} (\bar\nu_i\,\gamma_\mu P_L \nu_j) (\bar s\,\gamma^\mu d)
\;+\;{\rm h.c.}\,,
\label{eq:Kpinunu-op}
\eeq
where the summation is over different neutrino flavour states.\footnote{That is, neutrinos can be treated as massless with flavour eigen-states defined with respect to the charged lepton mass basis. Traditionally, these would be labeled as $i,j=e, \mu, \tau$, while for simplicity we retain the notation $i,j=1,2,3$ also for neutrino flavour states. } The dimensionful Wilson coefficient is a sum of SM and NP contributions,
\beq
\hat C_{ij}^{\nu\nu}=- \frac{(\lambda_c X^{\ell_i} +\lambda_t X_t)}{\Lambda_{\nu\nu, \text{SM}}^2} \delta_{ij} +\frac{C_{L;sd}^{\prime ij}}{\Lambda_\text{pre}^2},
\eeq
where for the SM contributions we introduced a one loop electroweak effective scale,
\beq
{\Lambda_{\nu\nu, \text{SM}}}=\Big(\frac{\sqrt 2}{ G_F}\frac{ \pi }{\alpha}\Big)^{1/2}\sin\theta_W\simeq 3.47\,\text{TeV}.
\eeq
with $G_F$ the Fermi constant, and $\sin \theta_W$ the sine of the weak mixing angle. The SM contributions also depend on products of relevant CKM matrix elements, 
\beq
\label{eq:lambda:CKM}
\lambda_i^\text{CKM}=V_{is}^* V_{id},
\eeq
 and on loop functions $X^{\ell_i}, X_t$, where $ X^e=X^\mu$, $r_X^{e/\tau}\equiv X^e/X^\tau\simeq 1.5$, and $X_t\simeq 1.46$ (explicit forms and more precise numerical values can be found in \cite{Buras:2015qea,Buras:2024ewl,Brod:2021hsj,Brod:2010hi,Buchalla:1995vs}). Note that the SM contributions are flavour diagonal, while the NP contributions have both diagonal and off-diagonal terms (cf. also \cref{eq:CL:ij}), 
\beq
\label{eq:CLprime:ij}
C_{L;sd}^{\prime ij}=\frac{1}{2}\Big( [C_{LL}^{\ell q}]_{ij;21} - [C_{LR}^{\ell q}]_{ij;12}\Big)
 \eeq
 
 In the presence of NP contributions the SM prediction for the $K^+\to \pi^+\nu\bar\nu$ branching ratio \cite{Brod:2021hsj}
 \beq
  \label{eq:Br:K+:pi+nunu:SM}
 \text{Br}(K^+\to \pi^+\nu\bar \nu)\big|_\text{SM}=\kappa_+ \biggr[\biggr(\frac{\Im\lambda_t^\text{CKM}}{\lambda^5} X_t\biggr)^2+\biggr(\frac{\Re \lambda_c^\text{CKM}}{\lambda}P_c+\frac{\Re\lambda_t^\text{CKM}}{\lambda^5}X_t\biggr)^2\biggr],
 \eeq
gets modified to 
  \beq
 \begin{split}
 \text{Br}(K^+\to \pi^+\nu\bar \nu)&=\text{Br}(K^+\to \pi^+\nu\bar \nu)\big|_\text{SM}-\frac{2}{3}\kappa_+\biggr(\frac{\Lambda_{\nu\nu,\text{SM}}}{\Lambda_\text{pre}}\biggr)^2\biggr[ \frac{\Re( \lambda_t^{\text{CKM}*} \Tr C_{L;sd}')}{\lambda^5\cdot \lambda^5 } X_t 
 \\
 &+\frac{\Re\lambda_c^\text{CKM}}{\lambda}\cdot \frac{\Re (\Tr_w C_{L;sd}')}{\lambda^5}P_c\biggr]+\frac{\kappa_+}{3}\biggr(\frac{\Lambda_{\nu\nu,\text{SM}}}{\Lambda_\text{pre}}\biggr)^4\sum_{ij}\biggr(\frac{\big|C_{L;sd}^{\prime ij}\big|}{\lambda^5}\biggr)^2.
 \end{split}
 \eeq
Here, $\kappa_+=0.5417(26)\cdot 10^{-10}$, $\lambda=0.2265(5)$, and $P_c=0.354(9)$~\cite{Brod:2021hsj}. The weighted trace in the above expression takes into account the differing contributions from electron, muon and tau lepton running in the SM loops involving charm quarks, 
 \beq
 \Tr_w C_{L;sd}'\equiv \frac{3}{1+2 r_X^{e/\tau}} \Big[ r_X^{e/\tau} (C_{L;sd}^{\prime 11}+C_{L;sd}^{\prime 22})+ C_{L;sd}^{\prime 33}\Big].
 \eeq
Note that while for simplicity we did not display the small QED and dimension-8 operator corrections in \cref{eq:Br:K+:pi+nunu:SM}, these were included in the numerical value for the SM branching ratio, $\text{Br}(K^+\to \pi^+\nu\bar \nu)\big|_\text{SM}=7.73(61)\times10^{-11}$ \cite{Brod:2021hsj} that we use. 

Similarly, the SM prediction for $\text{Br}(K_L\to \pi^0\nu\bar \nu)$ \cite{Brod:2021hsj}, 
\beq
\text{Br}(K_L\to \pi^0\nu\bar \nu)\big|_\text{SM}=\kappa_L r_{\epsilon_K} \biggr(\frac{\Im\lambda_t^\text{CKM}}{\lambda^5} X_t\biggr)^2,
\eeq
gets modified in the presence of two scalar preon model to
\beq
\begin{split}
\text{Br}(K_L\to \pi^0\nu\bar \nu)=\text{Br}(K_L\to \pi^0\nu\bar \nu)\big|_\text{SM} \biggr[1&-\frac{2}{3}\frac{\Im(\Tr C_{L;sd}')}{X_t\Im \lambda_t^\text{CKM}}\frac{\Lambda_{\nu\nu,\text{SM}}^2}{\Lambda_\text{pre}^2}
\\
+&\frac{1}{3} \sum_{ij}
\biggr(\frac{\Im C_{L;sd}^{\prime ij}} {X_t \Im\lambda_t^\text{CKM}} \frac{\Lambda_{\nu\nu,\text{SM}}^2}{\Lambda_\text{pre}^2} \biggr)^2\biggr],
\end{split}
\eeq
where $\text{Br}(K_L\to \pi^0\nu\bar \nu)\big|_\text{SM}=2.59(29)\times10^{-11}$ \cite{Brod:2021hsj}. 

For both $K^+ \to \pi^+\nu\bar \nu$ and $K_L\to \pi^0\nu\bar \nu$ the interference term between NP and the SM, which scales as $1/\Lambda_\text{pre}^2$ and involves $\nu$ and $\bar \nu$ of the same flavour, can have either sign, and thus can either reduce or enhance the branching ratios. Parametrically, the dominant contributions are from $C_L^{\prime 22},C_L^{\prime 33}\sim {\mathcal O}(\kappa^5)$, where in our benchmark $C_L^{\prime 33}$ is numerically enhanced and thus dominates the NP contribution by more than an order of magnitude. The flavour-off diagonal terms contribute only at ${\mathcal O}(1/\Lambda_\text{pre}^4)$, and always lead to an enhancement of the branching ratios, though numerically these contributions are subleading. 

The NA62 preliminary combined 2016--2024 result \cite{Chang:2026vvx,NA62:2024pjp}
$ 
{\rm Br}(K^+\to\pi^+\nu\bar\nu)=\left(9.6^{+1.9}_{-1.8}\right)\times10^{-11},
$
and the KOTO limit
${\rm Br}(K_L\to\pi^0\nu\bar\nu)<2.2\times10^{-9}$ (90\% C.L.)~\cite{KOTO:2024zbl} translate to the following bounds on $\Lambda_\text{pre}$ assuming our benchmark,
\begin{align}
K^+\to\pi^+\nu\bar\nu:& \qquad \Lambda_{\rm pre} > \bench{13}\,{\rm TeV} \,,
\label{eq:Kpinunu-Lpre-Kp}
\\
K_L\to\pi^0\nu\bar\nu: & \qquad \Lambda_{\rm pre} > \bench{7.4}\,{\rm TeV}\, (\bench{79}\,{\rm TeV}),
\label{eq:Kpinunu-Lpre-KL}
\end{align}where in the parenthesis we quote the bound that could be achieved if the $K_L\to\pi^0\nu\bar\nu$ branching ratio were measured at the SM value with 10\% error, the stated goal of KOTO program. In the benchmark we find ${\rm BR} /{\rm BR}_{\rm SM} \approx 1- 0.77\times\qty({15~{\rm TeV}}/{\Lambda_{\rm pre}})^2 + \qty({15~{\rm TeV}}/{\Lambda_{\rm pre}})^4$ for $K^+\rightarrow \pi^+\nu\bar{\nu}$ i.e., negative interference with the Standard Model, while for $K_L\rightarrow \pi^0\nu\bar{\nu}$ there is positive interference with the SM, ${\rm BR} /{\rm BR}_{\rm SM} \approx 1+ 1.2\times\qty({22~{\rm TeV}}/{\Lambda_{\rm pre}})^2 + \qty({22~{\rm TeV}}/{\Lambda_{\rm pre}})^4$

\paragraph{Short distance contribution to $K_S\rightarrow \mu^+ \mu^-$.}  

Another observable which may offer complementary sensitivity is $K_S\rightarrow \mu^+ \mu^-$. 
The relevant part of the $\Delta S=1$ semileptonic effective Lagrangian is due to (axial)$\otimes$(axial) currents \cite{Dery:2021mct,Dery:2021vql}, which receive the following new physics contributions, cf. \cref{eq:semileptonic:both,eq:KLemu-Leff-general},
\beq
\label{eq:DeltaS=1:KSmumu:NP}
\mathcal L_{\rm eff}^{\Delta S=1}
\supset \frac{\big(C_{R;sd}^{2 2}-C_{L;sd}^{22}\big)}{2 \Lambda_{\rm pre}^2} (\bar{\mu} \gamma_\alpha \gamma_5 \mu)\,
(\bar s \gamma^\alpha \gamma_5 d)+ {\rm h.c.},
\eeq
with $C_{L(R);sd}^{2 2}$ given in \cref{eq:CL:ij,eq:CR:ij}.
In the SM the short distance contributions are given by the effective Lagrangian \cite{Brod:2022khx}, 
\beq
\label{eq:DeltaS=1:KSmumu:SM}
{\cal L}_{\rm eff,SM}^{\Delta S=1}= \frac{G_F^2 m_W^2}{2 \pi^2} \Im\lambda_t^\text{CKM} Y_t \,(\bar{\mu} \gamma_\alpha \gamma_5 \mu)\,
(\bar s \gamma^\alpha \gamma_5 d)+ {\rm h.c.}+\cdots.
\eeq
where $\lambda_t^\text{CKM}$ is defined in \cref{eq:lambda:CKM}, $Y_t\simeq 0.931$ is the loop function, and we have only displayed the part of the short distance contributions that can be determined by measuring $\text{Br}(K_S\to \mu^+\mu^-)_{\ell=0}$ through time dependent analysis \cite{DAmbrosio:2017klp,Dery:2021mct,Dery:2021vql,Brod:2022khx,DAmbrosio:2025mxa}. At the end of Upgrade 2 LHCb could be able to exclude any NP contributions, \cref{eq:DeltaS=1:KSmumu:NP}, that exceed $35\%$ of the SM one \cite{DAmbrosio:2025mxa}, \cref{eq:DeltaS=1:KSmumu:SM}, leading to possible future constraint
\beq
\Lambda_\text{pre}>\bench{3.2}\,\text{TeV}\times\biggr(\frac{\Im(C_{R;sd}^{22}-C_{L;sd}^{22})}{\bench{4.1 \cdot 10^{-5}}}\biggr)^{1/2},
\eeq
where we used the benchmark values of $C_{L(R);sd}^{22}$.

\subsection{Four-quark operators}

\begin{figure}[t!]
  \begin{center}
 \includegraphics[width=0.6\textwidth]{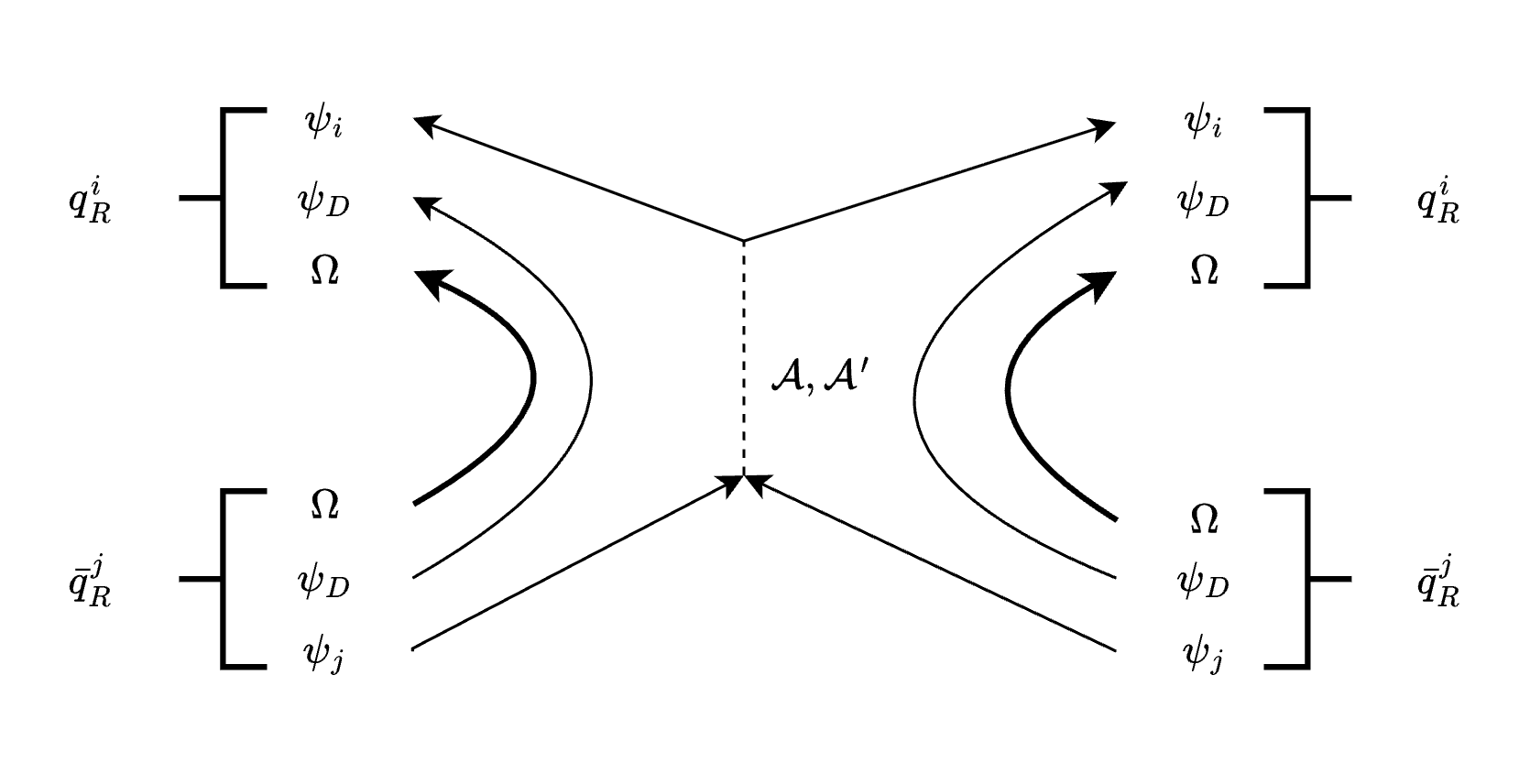}   
   \caption{Tree-level diagram for neutral meson mixing $M^0 \leftrightarrow \bar{M}^0$ from $\mathcal{A}$ and $\mathcal{A}'$ exchange. The diagram generates $\Delta F = 2$ operators with coefficient scaling as $|\lambda_{ij}^{(q)}|^4 / \Lambda_{\rm pre}^2$ at tree level. }
  \label{fig:meson_mixing_topology}
  \end{center}
\end{figure}

Four-quark operators are tightly constrained by neutral meson oscillations. In the SM, $M^0\leftrightarrow \bar{M}^0$ mixing arises at one loop through $W$-box diagrams \cite{Inami:1980fz,Aebischer:2020dsw}. In the two scalar preon model it is generated at tree level by $\mathcal{A}$ and $\mathcal{A}'$ exchange (Figure~\ref{fig:meson_mixing_topology}), leading to dimension-six $\Delta F = 2$ operators, cf. \cref{eq:semileptonic:both}
\begin{equation}
\label{eq:quark:mixing}
    \begin{split}
        \mathcal{L}_{\rm eff,NP}^{\Delta F=2} = \frac{1}{\Lambda_{\rm pre}^2}&\Big\{ [C_{LL}^{q q}]_{ij;ij} (\bar{q}_i \gamma^\mu P_L q_j)(\bar{q}_i \gamma_\mu P_L q_j) 
        \\
         & + [C_{RR}^{q q}]_{ji;ji} (\bar{q}_i \gamma^\mu P_R q_j)(\bar{q}_i \gamma_\mu P_R q_j) 
         \\
        &- \big([C_{LR}^{q q}]_{ij;ji} +[C_{RL}^{q q}]_{ji;ij}\big) (\bar{q}_i \gamma^\mu P_L q_j)(\bar{q}_i \gamma_\mu P_R q_j) \Big\}+ {\rm h.c.},
    \end{split}
\end{equation}
where the Wilson coefficients are given in \cref{eq:sum:CAB,eq:CAB:loop}, replacing $\ell\to q$. 
To interface with lattice matrix elements available in the literature we use the ``SUSY basis'' \cite{Gabbiani:1996hi} of chiral operators. Combinations involving $LR$ and $RL$ do not appear in the SUSY basis and must be rearranged with Fierz identities (superscripts denote colour). For example, denoting QCD-colour with $a$ and $b$, we have for $\Delta S=2$,
\begin{equation*}
    (\bar{s}^a\gamma_\mu P_L d_a)(\bar{s}^b \gamma^\mu P_R d_b)= -2\underbrace{(\bar{s}^a P_L d_b)(\bar{s}^b P_R d_a)}_{O_5^{sd}} ~.
\end{equation*}
In this basis the NP contributions to the effective $\Delta F=2$ Lagrangian are
\beq
 \mathcal{L}_{\rm eff,NP}^{\Delta F=2} = \frac{C_1^{q_iq_j}}{\Lambda_\text{pre}^2} O_1^{q_iq_j} + \frac{\tilde C_1^{q_iq_j}}{\Lambda_\text{pre}^2} \tilde O_1^{q_iq_j} +\frac{C_5^{q_iq_j}}{\Lambda_\text{pre}^2} O_5^{q_i q_j} +\cdots,
\eeq
where, $q=u,d$,
\begin{align}
C_1^{q_i q_j}&= [C_{LL}^{q q}]_{ij;ij}, 
\\
\tilde C_1^{q_i q_j}&= [C_{RR}^{q q}]_{ij;ij},
\\
C_5^{q_i q_j}&=2\big([C_{LR}^{q q}]_{ij;ji} +[C_{RL}^{q q}]_{ji;ij}\big).
\end{align}

 Lattice calculations show that the $O_5$ matrix element is roughly $10\times$ larger than that of $O_1$ and $\tilde O_1$ ($LL$ and $RR$ vector operators) \cite{Mescia:2012fg,Boyle:2012qb,ETM:2012vvy,Boyle:2024gge}. In deriving the bounds it thus suffices to the NP mixing contributions are dominated by $O_5$ operator (we have checked that this is the case for our numerical benchmark). Using the results by the UTfit collaboration on constraints on NP contributions to the meson mixing \cite{UTfit:2026qxi}, gives
 \begin{align}
 \label{eq:K-Kbar:Lambda}
 K-\bar K:& \qquad\Lambda_\text{pre}>\bench{8.5\cdot 10^3}\,\text{TeV} \times \biggr(\frac{|\Im C_5^{ds}|}{\bench{5.4\cdot 10^{-3}}}\biggr)^{1/2},
 \\
 D-\bar D:& \qquad\Lambda_\text{pre}>\bench{480}\,\text{TeV}\times \biggr(\frac{|\Im C_5^{uc}|}{\bench{1.0\cdot 10^{-3}}}\biggr)^{1/2},
\end{align}

For the $B$-systems the distinction between the real and imaginary parts is less pronounced and so we use the collaboration's result for the profiled likelihood \cite{UTfit:2026qxi} (this is a conservative choice), 
\begin{align}
\label{eq:B-Bbar:Lambda}
 B-\bar B:& \qquad \Lambda_\text{pre}>\bench{270}\,\text{TeV} \times \biggr(\frac{| C_5^{db}|}{\bench{1.2\cdot 10^{-2}}}\biggr)^{1/2},
  \\
  \label{eq:Bs-Bsbar:Lambda}
 B_s-\bar B_s:& \qquad \Lambda_\text{pre}>\bench{98}\,\text{TeV} \times \biggr(\frac{| C_5^{sb}|}{\bench{0.031}}\biggr)^{1/2}~.
 \end{align}
 In the future, an improved determination of $V_{cb}$ at FCC-ee~\cite{Charles:2020dfl}, together with an improved treatment of charm contributions~\cite{Brod:2026yxw,Brod:2021hsj,Brod:2019rzc}, can be expected to improve the $\Lambda_\text{pre}$ reach from $K$-$\bar K$ mixing in \cref{eq:K-Kbar:Lambda} by a factor of $\sim 2$, and for $B-\bar B$ and $B_s-\bar B_s$ a factor of $\sim 3\times$ improvement on $\Lambda_\text{pre}$ in \cref{eq:B-Bbar:Lambda,eq:Bs-Bsbar:Lambda} can be expected at the end of LHCb, Belle-II and FCC-ee programs \cite{Charles:2020dfl}.

\subsection{Leptonic four-fermion operators}

The tree-level and one loop exchanges of $\mathcal{A}$ and $\mathcal{A}'$ also generate purely leptonic four-fermion operators, resulting in the following effective Lagrangian (cf. \cref{eq:semileptonic:both} and \cref{fig:semileptonic_topology}),
\begin{equation}
\label{eq:leptonic}
    \begin{split}
        \mathcal{L}_{\rm lep} =\frac{1}{\Lambda_{\rm pre}^2} 
       & \Big\{ [C_{LL}^{\ell\ell}]_{ij;kl} 
        (\bar{\ell}_i \gamma^\mu P_L \ell_j)(\bar{\ell}_k \gamma_\mu P_L \ell_l) 
        \\
        &- [C_{LR}^{\ell\ell}]_{ij;lk} 
        (\bar{\ell}_i \gamma^\mu P_L \ell_j)(\bar{\ell}_k \gamma_\mu P_R \ell_l)
               \\
        &- [C_{RL}^{\ell\ell}]_{ji;kl} 
        (\bar{\ell}_i \gamma^\mu P_R \ell_j)(\bar{\ell}_k \gamma_\mu P_L \ell_l)
        \\
        &+ [C_{RR}^{\ell\ell}]_{ji;lk} 
        (\bar{\ell}_i \gamma^\mu P_R \ell_j)(\bar{\ell}_k \gamma_\mu P_R \ell_l)\Big\} + {\rm h.c.}, 
    \end{split}
\end{equation}
where $ [C_{AB}^{\ell\ell}]_{ij;kl} $ are the Wilson coefficients built out of the spurions $\lambda, \lambda'$ in the charged lepton mass basis. They are obtained by replacing $q\to \ell$ in \cref{eq:sum:CAB,eq:CAB:lq:tree,eq:CAB:loop}.

The above four-fermion operators mediate FCNC $\tau$ and $\mu$ decays such as $\mu\to 3e$ and $\tau \to 3\mu, 3e$, etc. However, these contributions are parametrically smaller than the contributions from the transition dipole moments, \cref{eq:Leff:dipole}. First of all, $v_d$ that enters the numerator in \cref{eq:Leff:dipole} is larger than both muon and tau mass, $v_d\gg m_{\mu, \tau}$. Furthermore also the dimensionless Wilson coefficients $[C_\gamma^\ell]_{ij}$, \cref{eq:Cf:parametric}, are parametrically larger than the combinations of $\lambda_{\ell_{L,R} \ell_{L,R}}$ in the leptonic four-fermion operators. For instance, these are the largest for $\tau\to 3\mu$ and are ${\mathcal O}(\kappa^4)$, to be compared with $[C_\gamma^\ell]_{23}\sim {\mathcal O}(\kappa^2) $. This large hierarchy also holds for the numerical benchmark, where $[C_\gamma^\ell]_{ij}$ are larger by at least one to two orders of magnitude than the corresponding dimensionless Wilson coefficients for the four-fermion operators. 

\subsection{Summary of flavour constraints}
\label{sec:results-discussion}
In \cref{fig:reach_summary} we present a summary of experimental sensitivities to $\Lambda_{\rm pre}$. The left panel shows the reach for each observable in the anarchic limit, i.e., assuming that
all flavour violating coefficients are ${\mathcal O}(1)$ (operationally, in previous subsections in the expressions for bounds on $\Lambda_\text{pre}$ we set the values of Wilson coefficients to unity and set $\cos\beta=1$). The right panel in \cref{fig:reach_summary} instead shows the same bounds, but for the benchmark values of couplings, i.e., it collects the bounds on $\Lambda_\text{pre}$ that we derived in previous subsections. 

Several qualitative features are worth highlighting. Under a flavour-anarchic prior, the electron EDM offers the highest sensitivity at $1.3\cdot 10^6$~TeV (and up to $3\cdot 10^8$~TeV in the future), while some of the other very sensitive flavour probes are $\mu$-$e$ conversion with projected sensitivity of $1.2\cdot 10^5$~TeV at Mu2e-II and COMET, $\epsilon_K$ at $2\cdot 10^5$~TeV after expected improvements on the theoretical predictions, the projected MEG-II $\mu\to e\gamma$ sensitivity at $6\cdot 10^4$~TeV, and the neutron EDM with projected future sensitivity at $4\cdot 10^4$\,TeV.

Once the benchmark flavour structure is imposed on the couplings, the ordering reshuffles. The electron EDM ($\sim \bench{9.5\cdot 10^3}$~TeV) still emerges as the strongest probe, with the projected $d_e\sim 10^{-34}\,e\,$cm future sensitivity translating to $\sim \bench{2\cdot 10^6}$~TeV. Interestingly, the latter is above the current bound on proton lifetime $\tau_p$ from Super-Kamiokande (grey bar), $\Lambda_{\rm pre}\sim 10^{5}$~TeV, obtained for central values of $C_8$ and the vectorlike-quark masses in \cref{eq:proton_bound} (cf.\ \cref{sec:proton}), and is even above the Hyper-Kamiokande projected reach. Note that $\Lambda_{\rm pre}\propto \tau_p^{1/8}$, so even an order-of-magnitude improvement in the proton lifetime shifts this bound only mildly, while bounds from flavour physics scale either as a square-root of the observable (for EDMs and meson mixing constraints), or with the fourth power (for branching ratios). For our benchmark the other currently most stringent constraints come from the CP-violating part of $K$-$\bar K$ mixing ($\sim \bench{8.5\cdot 10^3}$~TeV), $\mu\to e\gamma$ ($\sim \bench{9.0\cdot 10^2}$~TeV, MEG-II projection $\sim \bench{1.1\cdot 10^3}$~TeV), $D$-$\bar D$ mixing ($\sim \bench{480}$~TeV), $B_d$-$\bar B_d$ mixing ($\sim \bench{270}$~TeV), and the neutron EDM ($\sim \bench{240}$~TeV). The dipole-dominated $\mu\to e$ conversion currently sits at $\sim \bench{220}$~TeV and $\mu\to 3e$ at $\sim \bench{1.6\cdot 10^2}$~TeV, with Mu2e-II and Mu3e expected to increase the reach to $\sim \bench{2.5\cdot 10^3}$~TeV and $\sim \bench{1.6\cdot 10^3}$~TeV, respectively. The remaining four-fermion semileptonic operators ($K\to\pi\nu\bar\nu$, $K_L\to e\mu$, $K_S\to\mu\mu$) and the $\tau$ decays and $b\to s\gamma$ sit at or below a few tens of TeV in the benchmark due to the small off-diagonal couplings in the mass basis.

\begin{figure}[t!]
  \begin{center}
    \includegraphics[width=\textwidth]{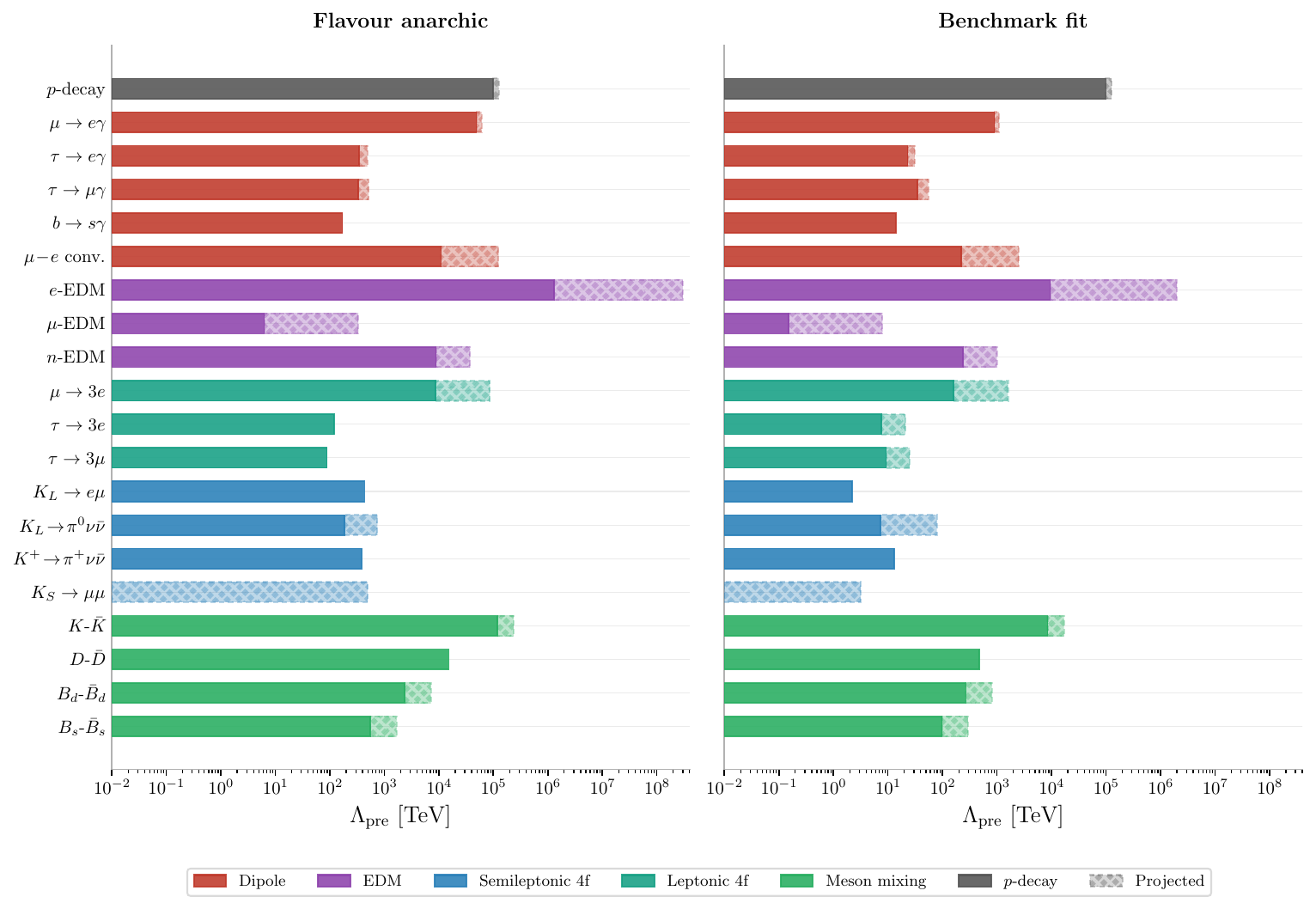}
    \caption{
    Experimental reach on the preonic scale $\Lambda_{\rm pre}$ from flavour and { CP}-violation observables, grouped by operator topology (proton decay, dipole and EDM operators, $\mu$-$e$ conversion, three-body leptonic decays, semileptonic four-fermion operators, and neutral-meson mixing). \textit{Left:} flavour-anarchic bounds, obtained by setting all flavour violating Wilson coefficients and $\cos\beta$ to unity. \textit{Right:} bounds evaluated with the benchmark values for $\lambda, \lambda'$ coupling matrices (see app. \ref{app:parameters}). For each observable the solid (dashed) bar shows the current (projected) experimental bounds. Note that in the bounds all $\SUp$ non-perturbative matrix elements were set to 1, while in general they can differ between, e.g., those for the muon-dipole operator and $K^0$--$\bar{K}^0$ oscillations; the sensitivity to $\Lambda_{\rm pre}$ scales roughly as the square root of these unknown parameters.}
    \label{fig:reach_summary}
  \end{center}
\end{figure}

The above ``benchmark hierarchy'' carries several caveats. First, in all the bounds we took the ${\mathcal O}(1)$ matrix elements, which capture the nonperturbative $\SUp$ dynamics, to be exactly 1. This expectation most probably holds only up to an order of magnitude or so --- for instance, when the related nonperturbative matrix elements were allowed to float in the fit to CKM and quark and lepton masses, the best fit values were found to have a hierarchy of about an order of magnitude (see app. \ref{app:parameters}). Which is ``the best'' observable to probe preon dynamics can thus easily change simply due to nonperturbative dynamics. 

Furthermore, the benchmark values of $\lambda, \lambda'$ spurions were assumed in our analysis to follow a particular hierarchical pattern, chosen because we anticipated the minimal amount of fine tuning required to reproduce the observed values of CKM and the masses of SM quarks and charged leptons. 
It is quite possible that different choices for the $\SUF$ breaking spurions, other than the ones taken in this manuscript, are realized in the UV. These different choices would generically lead to new structures for the lepton and quark Yukawas. As we have already seen in the two-scalar model given above, the lepton Yukawa will generically involve different spurions than the quark Yukawa. This can lead to flavour textures that differ substantially from either ``anarchy'' (left panel in \cref{fig:reach_summary}) or the benchmark (right panel).
It would be interesting to understand which features of the benchmark are ``generic'' in the $\SUp$ setup in Ref.~\cite{Dobrescu:2021fny}, and if other possibilities can be realized with different choices of spurions in the UV. For example one could break $\SUF$ by charging $\{\psi_i\}$ under a further additional gauge group. A similar question pertains to the size of CP violation in electron EDMs vs the CKM matrix.  

Finally, the bounds above apply to the operators generated at $\Lambda_{\rm pre}$. A state $X$ of mass $M_X$ contributes to a given observable as a correction to a dimension-6 Wilson coefficient of the size $C_X/M_X^2$, to be compared with $C/\Lambda_{\rm pre}^2$ from the local operator, and is subleading if $M_X/\Lambda_{\rm pre}\gtrsim (C_X/C)^{1/2}$. The heavy Higgs bosons of the type-II two-Higgs-doublet model have couplings aligned with the SM Yukawas and are constrained mainly by $b\to s\gamma$, $m_{H^\pm}\gtrsim 800$~GeV~\cite{Misiak:2020vlo}. The lightest pNGBs, those of isospin, couple derivatively to flavour-universal currents, so they induce no tree-level flavour-changing neutral currents and their couplings are suppressed by the fermion masses. The vectorlike fermions and the $\SUF/SU(3)_{\rm F}$ pNGBs couple to the SM fermions through the same spurions $\lambda$, $\lambda'$ as the local operators, and for them the condition above becomes a lower bound on their masses from the same observables. In the benchmark texture $\lambda_{24}$ and $\lambda_{34}$ are of order one, so the vectorlike fermions that contain a fourth-direction prebaryon $\Omega_{X4}$ couple to the second and third generations with order-one strength. With all order-one coefficients set to one, $\mu\to e\gamma$ and $\epsilon_K$ then suggest masses of order tens of TeV for the vectorlike leptons and hundreds of TeV for the vectorlike quarks, and the pNGB masses, of order $\langle\phi_{1/6}\rangle$, would be similar to the vectorlike quark masses. The vectorlike quarks entering \cref{eq:proton_bound} contain no such prebaryon and are not affected. These estimates carry the same uncertainties as the bounds on $\Lambda_{\rm pre}$, and lighter vectorlike fermions are allowed if the first two generations couple more weakly to the fourth direction.

In view of these ambiguities, the results shown in \cref{fig:reach_summary} should be interpreted rather conservatively: i) it is clear that flavour probes can be a competitive alternative to the searches for proton decay, and ii)
 it is valuable to have a full suite of flavour observables to probe different flavour combinations and scenarios. 

\section{Conclusions}
\label{sec:conclusions}
The idea that quarks and leptons could emerge from composite dynamics is compelling. When a composite Higgs emerges from the same confining dynamics there is a natural connection between the composite structure of the fermions and flavour physics. 

Recently a $\SUp$ confining chiral gauge theory has been proposed, whose low energy dynamics may plausibly generate the Standard Model. Until now, the flavour structure of the model has not been considered in detail. We have taken the first step in this direction by investigating both how quark and lepton Yukawas are generated, and correlated predictions for flavour physics processes. This basic connection has long been understood, however we find that by demanding realistic quark and lepton Yukawa matrices a surprising amount of freedom is removed from the model.

The quark Yukawa matrices receive both a tree-level contribution from $\mathcal{A}$ and $\mathcal{A}'$ exchange and a one-loop crossed-box dressing weighted by $N/(16\pi^{2})$, while the lepton Yukawa carries an overall $1/N$ suppression from the mismatch in preon species between the lepton prebaryons and the constituents of $H_d$. Hierarchical $\kappa$-textures in the flavour spurions $\lambda$ and $\lambda'$ naturally generate the observed mass hierarchies, with heavy-block entries of order unity and light-block entries suppressed by integer powers of $\kappa\approx 0.17$. The numerical fit reveals that all of the fitted non-perturbative coefficients are $O(1)$, and that the up/down splittings $\delta_X$ are bounded by $|\delta_X|\lesssim 1$. The first-generation observables ($m_u$, the small CKM elements, and the Jarlskog invariant $\mathcal{J}$) are the only places where some internal cancellations among the leading spurion-product structures are required, with the $m_u$ and $V_{us}$ adjustments essentially independent (Pearson $|r|\lesssim \bench{0.15}$).

Taking the proton-decay bound $\Lambda_{\rm pre}\sim 10^{5}$~TeV from \cref{sec:proton} as our reference, flavour physics offers a competitive and complementary probe of the preon scale. The strongest single bound in the benchmark is the electron EDM at $\sim \bench{9.5\cdot 10^{3}}$~TeV, whose projected $d_e\sim 10^{-34}\,e\,$cm sensitivity reaches $\sim \bench{2\cdot 10^{6}}$~TeV --- overtaking even the Hyper-Kamiokande proton-decay projection. Currently, similar sensitivity reach is obtained from the CP-violating part of $K$-$\bar K$ mixing ($\sim \bench{8.5\cdot 10^{3}}$~TeV for our benchmark), though here the projected future gain is expected to be more modest. The next tier comprises $\mu\to e\gamma$ ($\sim \bench{9.0\cdot 10^{2}}$~TeV), the neutron EDM ($\sim \bench{240}$~TeV), $D$-$\bar D$ mixing ($\sim \bench{480}$~TeV) and $B_d$-$\bar B_d$ mixing ($\sim \bench{270}$~TeV), with the dipole-dominated $\mu\to e$ conversion and $\mu\to 3e$ projected to reach $\sim \bench{2.5\cdot 10^{3}}$~TeV and $\sim \bench{1.6\cdot 10^{3}}$~TeV at Mu2e-II and Mu3e, respectively. Other flavour probes ($K\to\pi\nu\bar\nu$, $K_L\to e\mu$, $K_S\to\mu\mu$, $\tau$ decays, 
$b\to s(d)\gamma$) sit at the few to few-tens of TeV level in this benchmark.

The e-EDM is the most powerful probe under an anarchic flavour structure ($\sim 1.3\cdot 10^6$~TeV model-independently), but is reduced to the few$\times \bench{10^{3}}$~TeV level by the small { CP}-violating combination of lepton rotations and dipole flavour structures realized in the fit. Four-fermion semileptonic and leptonic operators are similarly suppressed by the small off-diagonal lepton couplings in the mass basis, reflecting the near-diagonal structure of the lepton Yukawa rotation. To probe the sensitivity of these conclusions to higher-order corrections we have included the crossed-box dressing of all four-fermion operators on the same footing as the tree exchange.

Our results have been presented as a probe of $\Lambda_{\rm pre}$, however it is possible that the theory hosts scalars that are parametrically lighter than $\Lambda_{\rm pre}$ such that the scale being probed is the scalar mass mediating the transition rather than $\Lambda_{\rm pre}$. These non-local contributions mediated by scalars lighter than $\Lambda_{\rm pre}$ may contribute both to proton decay and flavour physics. An estimate of these effects, and a better understanding of the model's scalar sector is desirable and would help sharpen the comparison between flavour physics and proton decay. 

Finally, let us note that we have only analysed one concrete set of spurions that break the $\SUF$ flavour symmetry. With this setup we found that non-perturbative Yukawa couplings to the flavour spurions are essential. 

Future directions would be to investigate other flavour spurions, and to also study the possibility of misalignment induced by certain composite scalar vevs. Of particular interest would be a possible UV-embedding where the hierarchy of the flavour spurions is naturally explained. It may also be of interest to relax the assumption of a fine-tuned Higgs boson emerging as a sexaquark-like state. Instead, one could imagine that the composite scalars with the same electroweak charges as the Higgs are heavy and do not acquire a vev. The Higgs would then have to emerge from a separate electroweak symmetry breaking sector that couples to the flavour-preons $\psi_i$ and simultaneously provides the necessary flavour-breaking spurions. Such a sector could be a second confining group near the TeV scale. The SM Yukawas would then come from couplings between the two sectors generated at $\Lambda_{\rm pre}$, which are suppressed by powers of the ratio of the two scales unless the operators involved have dimensions close to their free-field values~\cite{Luty:2004ye,Rattazzi:2008pe}. Low-scale supersymmetry would instead remove the quadratic sensitivity of the Higgs to $\Lambda_{\rm pre}$. With small supersymmetry breaking, however, the chiral symmetry of the preons is spontaneously broken~\cite{Csaki:2021aqv}, and a supersymmetric version of the model would have a different composite spectrum. The nonsupersymmetric theory, which large-$N$ arguments suggest confines with unbroken chiral symmetry~\cite{Karasik:2022gve}, can be separated from this regime by a phase transition~\cite{Assi:2025rjx}. 

In summary, we find that flavour physics offers a diverse basket of observables that can probe high-scale compositeness in the context of Dobrescu's model. In the benchmark studied, the most important probes are the electron EDM, $\epsilon_K$ in $K^0$--$\bar{K}^0$ mixing, and $\mu\to e\gamma$, with dipole-dominated $\mu\to e$ conversion and $\mu\to 3e$ becoming equally relevant in the future; the other sensitive probes include the neutron EDM, $D$-$\bar D$ mixing and $B_d$-$\bar B_d$ mixing.
Future work will investigate different flavour spurions and their ability to reproduce the observed quark and lepton Yukawas in the Standard Model.

\section*{Acknowledgments}
We thank Bogdan Dobrescu for many useful and pedagogical discussions about the model studied in this work. We also thank Kaladi Babu, Graham Kribs, Claudio Manzari, Matthew McCullough, Maxim Pospelov, and Mario Reig for insightful comments and suggestions. BA and JZ acknowledge support in part by DOE grant DE-SC0011784, DE-SC0026301, and by NSF grants OAC-2103889, OAC-2411215, and OAC-2417682. AS is partially supported by the Grant No. NSF PHY-2310363, OAC-2417682 and also by a QUP Fellowship.

\appendix

\section{Flavour spurions}
\label{app:sec:spurions}
 The set of spurions one obtains at low-energies depends on how $\SUF$ is broken. For example, had we taken $\mathcal{A}$ and $\mathcal{A}'$ to transform in other representations then the Yukawa couplings would have different symmetry properties. Alternatively, if $\SUF$ is broken by heavy vector fields (which may themselves be composite and descend from some prior stage of tumbling) then different flavour spurions are singled out because the chiral Feynman rules differ between scalars and vectors. Finally, gauge charge assignments that differ between the $\{\psi_i\}$ fields offers another possible mechanism for breaking $\SUF$

Let us consider the relevant $\SUF$ flavour building blocks in the theory. At the renormalizable level these are $\psi_i$, $\psi_{[i}\psi_{j]}$, and $\psi_{i} \psi_{j}$ where $[\ldots]$ denotes an antisymmetrized, and $\{ \ldots \}$ a symmetrized, pair of indices; we have the $\mathbf{4}$, the $\mathbf{6}$, and the $\mathbf{10}$ of $\SUF$ respectively. To couple to these objects and produce a singlet requires the conjugate representations. Single representations can be naturally picked out due to compatibility with Lorentz invariance and the $\SUp$ dynamics. 

In the realization adopted in the main text the two scalars sit in different $\SUp$ representations, $\mathcal A\in\overline{\mathbf{105}}$ (antisymmetric two-tensor) and $\mathcal A'\in\overline{\mathbf{120}}$ (symmetric two-tensor), and couple to pairs of fundamental preons through the Yukawa interactions
\begin{equation}
\mathcal{L}_{\rm Yuk} = \frac12\lambda_{ij}\,\mathcal A_{ab}\,\psi_i^a \psi_j^b
       + \frac12 \lambda'_{ij}\,\mathcal A'_{ab}\,\psi_i^a \psi_j^b + {\rm h.c.}\,,
\label{eq:Yukawa}
\end{equation}
where $i,j=1,\dots,4$ are preon flavour indices and $a,b=1,\dots,15$ are $\SUp$ colour indices. Contracting the $\SUp$ indices then forces $\lambda_{ij}$ to be antisymmetric (transforming in the $\overline{\mathbf{6}}$ of $\SUF$) and $\lambda'_{ij}$ to be symmetric (transforming in the $\overline{\mathbf{10}}$ of $\SUF$).

Because $\lambda'$ is complex symmetric it admits a Takagi factorization with real positive singular values, $\lambda'_{ij} = U^{T}\,{\rm diag}(\lambda'_1,\dots,\lambda'_4)\,U$. We use this freedom to choose a basis of the $\psi_i$ in which $\lambda'$ is real and diagonal; the antisymmetric matrix $\lambda$ then retains physical complex phases (its Youla normal form has $2\times 2$ complex blocks rather than a real diagonal). Two independent spurions are needed to generate a non-trivial CKM matrix: if a single scalar were responsible for $\SUF$ breaking it could always be brought to a flavour-aligned form, leaving the up- and down-type Yukawas proportional. The same conclusion holds had we instead taken both scalars in the $\overline{\mathbf{105}}$, in which case both spurions would be antisymmetric, or used a vector mediator, in which case the corresponding spurion can be diagonalized with standard unitaries.

\section{Benchmark values of parameters}
\label{app:parameters}

This appendix presents the complete set of fitted parameters from the numerical minimization described in Section~\ref{sec:masses}. Although the benchmark fit is illustrative (and should not be taken seriously beyond $O(1)$ agreement), it is important that the model has enough freedom to accommodate the CKM matrix and mass hierarchies. 

In the fit we set $N= 15$, giving a loop factor $N/(16\pi^{2}) \approx 0.1$ in the expression for the Yukawa matrices, \cref{eq:quark_yukawa,eq:lepton_yukawa}. The flavour texture expansion parameter was set to $\kappa^{2}=0.03$ (i.e., $\kappa\approx 0.17$).

The benchmark value of the ratio of $H_u$ and $H_d$ vevs is 
\begin{equation}
\tan\beta = 22.87, \qquad \Rightarrow \qquad \cos\beta=0.0437,
\end{equation}
while the two vevs are
\beq
v_u = v \sin\beta =245.77\text{ GeV},\qquad v_d = v\cos\beta= 10.75\text{ GeV}.
\eeq
The antisymmetric $\lambda$ and symmetric $\lambda'$ spurions take the following benchmark values (here and below the parameters are displayed to four decimal places; the complete benchmark -- spurions, rotation matrices, and non-perturbative coefficients -- is provided at full numerical precision in the machine-readable ancillary file \texttt{benchmark\_full\_precision.txt} accompanying the arXiv submission)
\begin{align}
\lambda &= \begin{pmatrix}
        +0.0000+0.0000\,i & -0.0010+0.0002\,i & +0.0026-0.0056\,i & -0.1841-0.1631\,i \\
        +0.0010-0.0002\,i & +0.0000+0.0000\,i & -0.0032-0.0164\,i & -1.2146+0.3235\,i \\
        -0.0026+0.0056\,i & +0.0032+0.0164\,i & +0.0000+0.0000\,i & +0.6939+0.9143\,i \\
        +0.1841+0.1631\,i & +1.2146-0.3235\,i & -0.6939-0.9143\,i & +0.0000+0.0000\,i \\
\end{pmatrix},
\\
\lambda'&= \begin{pmatrix}
        +0.0001-0.0008\,i & +0.0001-0.0012\,i & -0.0006-0.0038\,i & -0.0037-0.0033\,i \\
        +0.0001-0.0012\,i & -0.0030+0.0040\,i & +0.0495+0.0042\,i & +0.1430-0.0411\,i \\
        -0.0006-0.0038\,i & +0.0495+0.0042\,i & +1.4470-0.8030\,i & +0.3092+1.0487\,i \\
        -0.0037-0.0033\,i & +0.1430-0.0411\,i & +0.3092+1.0487\,i & -1.1534-0.0113\,i \\
\end{pmatrix},
\end{align}
where the antisymmetry of $\lambda$ sets the diagonal elements to zero, $\lambda_{aa}=0$. Note that the benchmark spurions obey the parametric scaling with $\kappa$ in \cref{eq:lambda_texture}.

The singular value decomposition of Yukawa matrices, $Y_f = U_f \, \mathrm{diag}(y_1,y_2,y_3) \, V_f^{\dagger}$, cf. \cref{eq:Yud:rotate,eq:Yell:rotate} defines the left ($U_f$) and right ($V_f$) rotation matrices for $f=u,d,\ell$. 
For up-quark sector these are
\begin{align}
U_u &= \begin{pmatrix}
        \bench{-0.6448-0.4835\,i} & \bench{+0.5871+0.0756\,i} & \bench{+0.0038-0.0020\,i} \\
        \bench{+0.2704+0.5265\,i} & \bench{+0.6737+0.4422\,i} & \bench{-0.0137+0.0133\,i} \\
        \bench{-0.0006-0.0140\,i} & \bench{-0.0036-0.0131\,i} & \bench{-0.9847+0.1731\,i} \\
\end{pmatrix},
\\
V_u &= \begin{pmatrix}
        \bench{-0.8406-0.5233\,i} & \bench{-0.0840-0.1112\,i} & \bench{+0.0062+0.0040\,i} \\
        \bench{+0.0007+0.1395\,i} & \bench{+0.3517-0.9252\,i} & \bench{+0.0186+0.0236\,i} \\
        \bench{+0.0040+0.0027\,i} & \bench{-0.0182+0.0245\,i} & \bench{+0.3627+0.9314\,i} \\
\end{pmatrix},
\end{align}
and for down-quark sector
\begin{align}
U_d &= \begin{pmatrix}
        \bench{-0.7612-0.4881\,i} & \bench{+0.4246-0.0362\,i} & \bench{+0.0273+0.0029\,i} \\
        \bench{+0.1105+0.4124\,i} & \bench{+0.7174+0.5491\,i} & \bench{+0.0189+0.0330\,i} \\
        \bench{-0.0076-0.0057\,i} & \bench{+0.0399-0.0228\,i} & \bench{-0.9839+0.1724\,i} \\
\end{pmatrix},
\\
V_d & = \begin{pmatrix}
        \bench{-0.9901+0.0000\,i} & \bench{-0.1339-0.0239\,i} & \bench{+0.0242+0.0241\,i} \\
        \bench{+0.0110+0.1340\,i} & \bench{+0.1149-0.9800\,i} & \bench{+0.0850+0.0327\,i} \\
        \bench{+0.0386+0.0095\,i} & \bench{-0.0757+0.0463\,i} & \bench{+0.2447+0.9647\,i} \\
\end{pmatrix}.
\end{align}
The column-phase convention for $U_u$ and $U_d$ above is chosen so that $\hat V_{\rm CKM} = U_u^{\dagger} U_d$ comes out in the PDG standard parametrization; the corresponding column phases of $V_u$, $V_d$ are inherited.
In terms of powers of the expansion parameter $\kappa$, taking as the dividing lines $\sqrt{1/\kappa}=2.42$ and $\sqrt{\kappa}=0.41$, this gives
\beq
U_u\sim 
\begin{pmatrix}
  1 & 1 & \kappa^{3}
  \\
 1 & 1 & \kappa^2
 \\
 \kappa^{2} & \kappa^2 & 1
 \end{pmatrix}, 
\,
V_u\sim 
\begin{pmatrix}
  1 & \kappa & \kappa^{3}
  \\
 \kappa & 1 & \kappa^2
 \\
 \kappa^{3} & \kappa^2 & 1
 \end{pmatrix}, 
\,
 U_d\sim
 \begin{pmatrix}
  1 & 1 & \kappa^{2}
  \\
1& 1 & \kappa^2
 \\
 \kappa^{3} & \kappa^2 & 1
 \end{pmatrix},
\,
  V_d\sim 
 \begin{pmatrix}
  1 & \kappa & \kappa^{2}
  \\
 \kappa & 1 & \kappa
 \\
 \kappa^{2} & \kappa & 1
 \end{pmatrix}.
\end{equation}
For charged leptons the two SVD rotation matrices are
\begin{align}
U_\ell & = \begin{pmatrix}
        +0.6256+0.7761\,i & -0.0337-0.0539\,i & -0.0209+0.0432\,i \\
        -0.0375-0.0548\,i & -0.4650-0.8808\,i & -0.0349+0.0479\,i \\
        -0.0384-0.0216\,i & +0.0539+0.0310\,i & -0.0615+0.9952\,i \\
\end{pmatrix},
\\
V_\ell & = \begin{pmatrix}
        +0.9520+0.0000\,i & -0.3051+0.0000\,i & -0.0244+0.0000\,i \\
        -0.2171+0.2151\,i & -0.6759+0.6631\,i & -0.0204+0.0983\,i \\
        -0.0066-0.0159\,i & -0.0976-0.0294\,i & +0.9622-0.2521\,i \\
\end{pmatrix}.
\end{align}
In terms of the expansion parameter these are 
\beq
U_{\ell}\sim 
\begin{pmatrix}
  1 & \kappa^2 & \kappa^{2}
  \\
 \kappa^2 & 1 & \kappa^{2}
 \\
 \kappa^{2} & \kappa^{2} & 1
 \end{pmatrix}, 
 \qquad
V_{\ell}\sim
 \begin{pmatrix}
  1 & \kappa & \kappa^{2}
  \\
 \kappa & 1 & \kappa^{}
 \\
 \kappa^{2} & \kappa^{} & 1
 \end{pmatrix}.
\eeq

That is, the benchmark numerical values for the mixing matrices are mostly close to the naive expectation, 
\beq
\label{eq:naive:U:V}
U_{u,d,\ell}\sim V_{u,d,\ell}\sim V_\text{CKM},
\eeq
 with the deviations from this parametric expectation at most a factor of few (with the exception of $[U_\ell]_{13}, [U_\ell]_{31}$, which are almost an order of magnitude larger than the naive expectation). This motivates the use of \cref{eq:naive:U:V} in the estimates of generic expectations for the sizes of FCNCs, as we do in the main text.

\subsection{Non-perturbative coefficients/matrix elements}
\label{app:NDA-fit}
As emphasized in the main text we parametrize the impact of non-perturbative matrix elements in terms of coefficients which we demand to be $\sim O(1)$ when attempting to fit the observed quark and lepton Yukawa structure. 

The six non-perturbative matrix elements in the quark Yukawa $(F', F, G, I, J, K)$ (tree + box) take the fitted values
\begin{equation}
F' = \bench{+0.414},\; F = \bench{+0.684},\; G = \bench{+2.447},\; I = \bench{-0.432},\; J = \bench{-2.239},\; K = \bench{+1.871}.
\end{equation}
The analogous lepton-sector matrix elements are
\begin{equation}
F'^{\ell} = \bench{+0.404},\; F^{\ell} = \bench{-1.588},\; G^{\ell} = \bench{+0.000},\; I^{\ell} = \bench{-1.702},\; J^{\ell} = \bench{+5.647},\; K^{\ell} = \bench{-4.248}.
\end{equation}
The up/down vev-driven splittings $X^{d} = X^{u}(1 + \delta_X)$ with $|\delta_X| \le 1$ are
\begin{equation}
\begin{aligned}
\delta_{F'} = -0.685, \;
\delta_{F} = +0.962, \;
\delta_{G} = +1.000, \;
\delta_{I} = +0.262, \;
\delta_{J} = +0.913, \;
\delta_{K} = -0.823. \;
\end{aligned}
\end{equation}

\subsection{Residual fine-tuning}
\label{app:fine-tuning}

All tree-level non-perturbative matrix elements above lie within $|X|\lesssim 1.6$, while the box-topology coefficients satisfy $|X|\lesssim 6$; the $u/d$ splittings all obey $|\delta_X|\le 1$. The largest box coefficients are $J^{\ell}\simeq \bench{+5.6}$ and $G^d \simeq 2\, G^u\simeq \bench{4.9}$ (from $\delta_G\simeq +1$), followed by $J^{d}\simeq \bench{-4.3}$ and $K^{\ell}\simeq \bench{-4.2}$ -- all within a factor of $\lesssim 2$ of the $\mathcal O(1)$ NDA expectation. One channel collapses to zero, $G^{\ell}\simeq 0$. The $\kappa$-textures of \cref{eq:lambda_texture} are reproduced by amplitudes $|a_{ij}|\in[\bench{0.6},\bench{1.7}]$ at every entry, and the fit is ``natural''.

To probe robustness more sharply we have randomised all fitted parameters by a $3\%$ relative jitter and traced the induced shifts in each predicted observable (200 trials). The heavy-flavour observables $m_t$, $m_b$, $m_\tau$, the strange and charm quark masses, the muon mass, the diagonal CKM entries and $|V_{cb}|$, $|V_{ts}|$ all remain stable -- their relative shifts are commensurate with the parameter perturbation (amplification factors of order unity). The Cabibbo angle $|V_{us}|$ shows a moderate few-times amplification. Five observables exhibit large multiplicative response, dominated by internal cancellations: the first-generation Yukawa eigenvalues $m_u$ and $m_e$, the smallest CKM elements $|V_{ub}|$ and $|V_{td}|$, and the Jarlskog invariant $\mathcal{J}$. This residual sensitivity is structural. The small size of these observables in nature relative to the $\kappa$-scaling that one would ``naturally'' predict from our imposed textures demands a certain level of fine-tuning between the tree-level and box contributions in the first-generation sector. When one demands $\mathcal O(1)$ values for the non-perturbative coefficients the fit arranges for cancellations among the leading spurion-product structures. The model therefore reproduces the entire SM Yukawa pattern with a localised ``hot-spot'' of fine-tuning, sitting precisely where the deepest hierarchy of the SM flavour spectrum lies.

\section{Numerical minimization procedure}
\label{app:minimization}

We determine the model parameters by minimizing a loss function that measures the deviation between predicted and observed masses and CKM elements. The free parameters are the antisymmetric spurion $\lambda$ (six independent complex entries), the symmetric spurion $\lambda'$ (ten independent complex entries), $\tan\beta$, the six real non-perturbative coefficients $(F',F,G,I,J,K)$ in the quark Yukawa of \cref{eq:quark_yukawa} and their lepton-sector counterparts $(F'^{\ell},F^{\ell},G^{\ell},I^{\ell},J^{\ell},K^{\ell})$, and the six up/down splittings $\delta_{X}$ defined by $X^{d}=X^{u}(1+\delta_X)$. After fixing the residual $\SUF$ flavour redefinitions used to align $\lambda'$ this leaves on the order of $40$ real parameters that are varied in the fit.

To impose the $\kappa$-textures of \cref{eq:lambda_texture} we write each spurion entry as $a_{ij}\,\kappa^{n_{ij}}$ with $\kappa\approx 0.17$ and the integer powers $n_{ij}$ given there. The order-unity amplitudes $a_{ij}$ are fitted directly. The six non-perturbative coefficients in each sector are required to satisfy $|X|\lesssim 6$, in line with NDA estimates, while the six splittings are bounded by $|\delta_{X}|\le 1$; both bounds are imposed via soft barriers that vanish inside the allowed range and grow rapidly outside.

The loss function is a weighted sum of squared relative errors on the nine charged-fermion masses, the nine moduli of the CKM matrix, and the Jarlskog invariant,
\begin{equation}
\begin{split}
\mathcal{L} = \sum_{i\in\{q,\ell\}} &w_{m}\left(\frac{m_i^{\rm pred} - m_i^{\rm obs}}{m_i^{\rm obs}}\right)^2 \\
&+ w_{|V|}\sum_{jk}\!\left(|V_{jk}|^{\rm pred} - |V_{jk}|^{\rm obs}\right)^2 + w_J\!\left(\mathcal{J}^{\rm pred} - \mathcal{J}^{\rm obs}\right)^2 + \Delta_{\rm reg}\,,
\end{split}
\end{equation}
with empirical $\mathcal{O}(10^1$--$10^2)$ weights $w_m$, $w_{|V|}$, $w_J$ (mass residuals are evaluated logarithmically). Inputs are evolved to $\mu=10^{4}$~TeV using one-loop SM running. The regularization $\Delta_{\rm reg}$ collects the soft-barrier penalties on the fitted coefficients and splittings together with a mild push that keeps every spurion amplitude $|a_{ij}|\gtrsim 0.1$. We minimize using a sequence of Adam and L-BFGS phases on the unconstrained parametrization, with multiple random restarts.

\subsection{Fit quality}

Convergence is reached within a few thousand epochs. The fitted Higgs parameters are $\tan\beta=22.87$, $v_u=245.77$~GeV, $v_d=10.75$~GeV, and the predicted masses match their input values to better than $0.1\%$ both for the charged-fermion spectrum and for all nine $|V_{\rm CKM}|$ entries; the Jarlskog invariant is reproduced to comparable accuracy. The complete set of fitted spurions, matrix-element coefficients, splittings, and the resulting rotation matrices are tabulated in \cref{app:parameters}; they form the input for the flavour-violating analysis of \cref{sec:flavour}. The robustness of this benchmark to small parameter perturbations is discussed in \cref{app:fine-tuning}. Since we have not considered higher order corrections agreement to $\sim 10\%$ is, in fact, sufficient. The sub-percent agreement should be taken as evidence that small adjustments in the matrix elements can accommodate any higher order corrections that we have not explicitly accounted for. 

\bibliographystyle{JHEP}
\bibliography{preons_bibl}

\end{document}